
\magnification=\magstep1
\vsize=46.5pc
%


\catcode`\@=11


\message{Loading a modification of the jyTeX macros...}

\message{modifications to plain.tex,}


\def\newcount{\alloc@0\count\countdef\insc@unt}
\def\newdimen{\alloc@1\dimen\dimendef\insc@unt}
\def\newskip{\alloc@2\skip\skipdef\insc@unt}
\def\newmuskip{\alloc@3\muskip\muskipdef\@cclvi}
\def\newtoks{\alloc@5\toks\toksdef\@cclvi}
\def\newhelp#1#2{\newtoks#1\global#1\expandafter{\csname#2\endcsname}}
\def\newread{\alloc@6\read\chardef\sixt@@n}
\def\newwrite{\alloc@7\write\chardef\sixt@@n}
\def\newfam{\alloc@8\fam\chardef\sixt@@n}
\def\newinsert#1{\global\advance\insc@unt by\m@ne
     \ch@ck0\insc@unt\count
     \ch@ck1\insc@unt\dimen
     \ch@ck2\insc@unt\skip
     \ch@ck4\insc@unt\box
     \allocationnumber=\insc@unt
     \global\chardef#1=\allocationnumber
     \wlog{\string#1=\string\insert\the\allocationnumber}}
\def\newif#1{\count@\escapechar \escapechar\m@ne
     \expandafter\expandafter\expandafter
          \xdef\@if#1{true}{\let\noexpand#1=\noexpand\iftrue}%
     \expandafter\expandafter\expandafter
          \xdef\@if#1{false}{\let\noexpand#1=\noexpand\iffalse}%
     \global\@if#1{false}\escapechar=\count@}


\newlinechar=`\^^J
\overfullrule=0pt

\message{hacks,}


\toksdef\toks@i=1
\toksdef\toks@ii=2


\def\TeX{T\kern-.1667em \lower.5ex \hbox{E}\kern-.125em X\null}
\def\jyTeX{{\leavevmode
     \raise.587ex \hbox{\it\j}\kern-.1em \lower.048ex \hbox{\it y}\kern-.12em
     \TeX}}

\let\then=\iftrue
\def\ifnoarg#1\then{\def\hack@{#1}\ifx\hack@\empty}
\def\ifundefined#1\then{%
     \expandafter\ifx\csname\expandafter\blank\string#1\endcsname\relax}
\def\useif#1\then{\csname#1\endcsname}
\def\usename#1{\csname#1\endcsname}
\def\useafter#1#2{\expandafter#1\csname#2\endcsname}

\long\def\loop#1\repeat{\def\@iterate{#1\expandafter\@iterate\fi}\@iterate
     \let\@iterate=\relax}

\let\TeXend=\end
\def\begin#1{\begingroup\def\@@blockname{#1}\usename{begin#1}}
\def\End#1{\usename{end#1}\def\hack@{#1}%
     \ifx\@@blockname\hack@
          \endgroup
     \else\err@badgroup\hack@\@@blockname
     \fi}
\def\@@blockname{}

\def\defaultoption[#1]#2{%
     \def\hack@{\ifx\hack@ii[\toks@={#2}\else\toks@={#2[#1]}\fi\the\toks@}%
     \futurelet\hack@ii\hack@}

\def\markup#1{\let\@@marksf=\empty
     \ifhmode\edef\@@marksf{\spacefactor=\the\spacefactor\relax}\/\fi
     ${}^{\hbox{\subscriptfonts#1}}$\@@marksf}


\newtoks\shortyear
\newtoks\militaryhour
\newtoks\standardhour
\newtoks\minute
\newtoks\amorpm

\def\settime{\count@=\time\divide\count@ by60
     \militaryhour=\expandafter{\number\count@}%
     {\multiply\count@ by-60 \advance\count@ by\time
          \xdef\hack@{\ifnum\count@<10 0\fi\number\count@}}%
     \minute=\expandafter{\hack@}%
     \ifnum\count@<12
          \amorpm={am}
     \else\amorpm={pm}
          \ifnum\count@>12 \advance\count@ by-12 \fi
     \fi
     \standardhour=\expandafter{\number\count@}%
     \def\hack@19##1##2{\shortyear={##1##2}}%
          \expandafter\hack@\the\year}

\def\monthword#1{%
     \ifcase#1
          $\bullet$\err@badcountervalue{monthword}%
          \or January\or February\or March\or April\or May\or June%
          \or July\or August\or September\or October\or November\or December%
     \else$\bullet$\err@badcountervalue{monthword}%
     \fi}

\def\monthabbr#1{%
     \ifcase#1
          $\bullet$\err@badcountervalue{monthabbr}%
          \or Jan\or Feb\or Mar\or Apr\or May\or Jun%
          \or Jul\or Aug\or Sep\or Oct\or Nov\or Dec%
     \else$\bullet$\err@badcountervalue{monthabbr}%
     \fi}

\def\militarytime{\the\militaryhour:\the\minute}
\def\standardtime{\the\standardhour:\the\minute}


\def\@setnumstyle#1#2{\expandafter\global\expandafter\expandafter
     \expandafter\let\expandafter\expandafter
     \csname @\expandafter\blank\string#1style\endcsname
     \csname#2\endcsname}
\def\numstyle#1{\usename{@\expandafter\blank\string#1style}#1}
\def\ifblank#1\then{\useafter\ifx{@\expandafter\blank\string#1}\blank}

\def\blank#1{}

\def\Roman#1{\expandafter\uppercase\expandafter{\romannumeral#1}}
\def\alphabetic#1{%
     \ifcase#1
          $\bullet$\err@badcountervalue{alphabetic}%
          \or a\or b\or c\or d\or e\or f\or g\or h\or i\or j\or k\or l\or m%
          \or n\or o\or p\or q\or r\or s\or t\or u\or v\or w\or x\or y\or z%
     \else$\bullet$\err@badcountervalue{alphabetic}%
     \fi}
\def\Alphabetic#1{\expandafter\uppercase\expandafter{\alphabetic{#1}}}
\def\symbols#1{%
     \ifcase#1
          $\bullet$\err@badcountervalue{symbols}%
          \or*\or\dag\or\ddag\or\S\or$\|$%
          \or**\or\dag\dag\or\ddag\ddag\or\S\S\or$\|\|$%
     \else$\bullet$\err@badcountervalue{symbols}%
     \fi}


\catcode`\^^?=13 \def^^?{\relax}

\def\trimleading#1\to#2{\edef#2{#1}%
     \expandafter\@trimleading\expandafter#2#2^^?^^?}
\def\@trimleading#1#2#3^^?{\ifx#2^^?\def#1{}\else\def#1{#2#3}\fi}

\def\trimtrailing#1\to#2{\edef#2{#1}%
     \expandafter\@trimtrailing\expandafter#2#2^^? ^^?\relax}
\def\@trimtrailing#1#2 ^^?#3{\ifx#3\relax\toks@={}%
     \else\def#1{#2}\toks@={\trimtrailing#1\to#1}\fi
     \the\toks@}

\def\trim#1\to#2{\trimleading#1\to#2\trimtrailing#2\to#2}

\catcode`\^^?=15


\long\def\additemL#1\to#2{\toks@={\^^\{#1}}\toks@ii=\expandafter{#2}%
     \xdef#2{\the\toks@\the\toks@ii}}

\long\def\additemR#1\to#2{\toks@={\^^\{#1}}\toks@ii=\expandafter{#2}%
     \xdef#2{\the\toks@ii\the\toks@}}

\def\getitemL#1\to#2{\expandafter\@getitemL#1\hack@#1#2}
\def\@getitemL\^^\#1#2\hack@#3#4{\def#4{#1}\def#3{#2}}


\newskip\headskip
\newskip\footskip

\message{document layout,}

\newif\ifdraft
\def\draft{\drafttrue\leftmargin=.5in \overfullrule=5pt }


\newskip\abovechapterskip
\newskip\belowchapterskip
\newskip\abovesectionskip
\newskip\belowsectionskip
\newskip\abovesubsectionskip
\newskip\belowsubsectionskip

\def\chapterstyle#1{\global\expandafter\let\expandafter\@chapterstyle
     \csname#1text\endcsname}
\def\sectionstyle#1{\global\expandafter\let\expandafter\@sectionstyle
     \csname#1text\endcsname}
\def\subsectionstyle#1{\global\expandafter\let\expandafter\@subsectionstyle
     \csname#1text\endcsname}

\def\CHapter#1{%
     \ifdim\lastskip=17sp \else\chapterbreak\vskip\abovechapterskip\fi
     \@chapterstyle{\ifblank\chapternumstyle\then
          \else\newchapternum=\next\chapternumformat\ \fi#1}%
     \nobreak\vskip\belowchapterskip\vskip17sp }

\def\Section#1{%
     \ifdim\lastskip=17sp \else\sectionbreak\vskip\abovesectionskip\fi
     \@sectionstyle{\ifblank\sectionnumstyle\then
          \else\newsectionnum=\next\sectionnumformat\ \fi#1}%
     \nobreak\vskip\belowsectionskip\vskip17sp }

\def\subsection#1{%
     \ifdim\lastskip=17sp \else\subsectionbreak\vskip\abovesubsectionskip\fi
     \@subsectionstyle{\ifblank\subsectionnumstyle\then
          \else\newsubsectionnum=\next\subsectionnumformat\ \fi#1}%
     \nobreak\vskip\belowsubsectionskip\vskip17sp }


\newtoks\everybye \everybye={\par\vfil}
\outer\def\bye{\the\everybye
     \footnotecheck
     \prelabelcheck
     \streamcheck
     \supereject
     \TeXend}

\message{labels,}

\let\@@labeldef=\xdef
\newif\if@labelfile
\newwrite\@labelfile
\let\@prelabellist=\empty

\def\Label#1#2{\trim#1\to\@@labarg\edef\@@labtext{#2}%
     \edef\@@labname{lab@\@@labarg}%
     \useafter\ifundefined\@@labname\then\else\@yeslab\fi
     \useafter\@@labeldef\@@labname{#2}%
     \ifstreaming
          \expandafter\toks@\expandafter\expandafter\expandafter
               {\csname\@@labname\endcsname}%
          \immediate\write\streamout{\noexpand\Label{\@@labarg}{\the\toks@}}%
     \fi}
\def\@yeslab{%
     \useafter\ifundefined{if\@@labname}\then
          \err@labelredef\@@labarg
     \else\useif{if\@@labname}\then
               \err@labelredef\@@labarg
          \else\global\usename{\@@labname true}%
               \useafter\ifundefined{pre\@@labname}\then
               \else\useafter\ifx{pre\@@labname}\@@labtext
                    \else\err@badlabelmatch\@@labarg
                    \fi
               \fi
               \if@labelfile
               \else\global\@labelfiletrue
                    \immediate\write\sixt@@n{--> Creating file \jobname.lab}%
                    \immediate\openout\@labelfile=\jobname.lab
               \fi
               \immediate\write\@labelfile
                    {\noexpand\prelabel{\@@labarg}{\@@labtext}}%
          \fi
     \fi}

\def\putlab#1{\trim#1\to\@@labarg\edef\@@labname{lab@\@@labarg}%
     \useafter\ifundefined\@@labname\then\@nolab\else\usename\@@labname\fi}
\def\@nolab{%
     \useafter\ifundefined{pre\@@labname}\then
          \undefinedlabelformat
          \err@needlabel\@@labarg
          \useafter\xdef\@@labname{\undefinedlabelformat}%
     \else\usename{pre\@@labname}%
          \useafter\xdef\@@labname{\usename{pre\@@labname}}%
     \fi
     \useafter\newif{if\@@labname}%
     \expandafter\additemR\@@labarg\to\@prelabellist}

\def\prelabel#1{\useafter\gdef{prelab@#1}}

\def\ifundefinedlabel#1\then{%
     \expandafter\ifx\csname lab@#1\endcsname\relax}
\def\useiflab#1\then{\csname iflab@#1\endcsname}

\def\prelabelcheck{{%
     \def\^^\##1{\useiflab{##1}\then\else\err@undefinedlabel{##1}\fi}%
     \@prelabellist}}

\message{equation numbering,}

\newcount\chapternum
\newcount\sectionnum
\newcount\subsectionnum
\newcount\equationnum
\newcount\subequationnum
\newcount\figurenum
\newcount\subfigurenum
\newcount\tablenum
\newcount\subtablenum
\newcount\defnum
\newcount\subdefnum
\newcount\thmnum
\newcount\subthmnum
\newcount\lemnum
\newcount\sublemnum

\newif\if@subeqncount
\newif\if@subfigcount
\newif\if@subtblcount
\newif\if@subdefcount
\newif\if@subthmcount
\newif\if@sublemcount

\def\newchapternum{\newsectionnum=\z@\@resetnum\chapternum}
\def\newsectionnum{\newsubsectionnum=\z@\@resetnum\sectionnum}
\def\newsubsectionnum{\newequationnum=\z@\newfigurenum=\z@\newtablenum=\z@
     \newdefnum=\z@\newthmnum=\z@\newlemnum=\z@
     \@resetnum\subsectionnum}
\def\newequationnum{\newsubequationnum=\z@\@resetnum\equationnum}
\def\newsubequationnum{\@resetnum\subequationnum}
\def\newfigurenum{\newsubfigurenum=\z@\@resetnum\figurenum}
\def\newsubfigurenum{\@resetnum\subfigurenum}
\def\newtablenum{\newsubtablenum=\z@\@resetnum\tablenum}
\def\newsubtablenum{\@resetnum\subtablenum}
\def\newdefnum{\newsubdefnum=\z@\@resetnum\defnum}
\def\newsubdefnum{\@resetnum\subdefnum}
\def\newthmnum{\newsubthmnum=\z@\@resetnum\thmnum}
\def\newsubthmnum{\@resetnum\subthmnum}
\def\newlemnum{\newsublemnum=\z@\@resetnum\lemnum}
\def\newsublemnum{\@resetnum\sublemnum}

\def\@resetnum#1{\global\advance#1by1 \edef\next{\the#1\relax}\global#1}

\newchapternum=0

\def\chapternumstyle#1{\@setnumstyle\chapternum{#1}}
\def\sectionnumstyle#1{\@setnumstyle\sectionnum{#1}}
\def\subsectionnumstyle#1{\@setnumstyle\subsectionnum{#1}}
\def\equationnumstyle#1{\@setnumstyle\equationnum{#1}}
\def\subequationnumstyle#1{\@setnumstyle\subequationnum{#1}%
     \ifblank\subequationnumstyle\then\global\@subeqncountfalse\fi
     \ignorespaces}
\def\figurenumstyle#1{\@setnumstyle\figurenum{#1}}
\def\subfigurenumstyle#1{\@setnumstyle\subfigurenum{#1}%
     \ifblank\subfigurenumstyle\then\global\@subfigcountfalse\fi
     \ignorespaces}
\def\tablenumstyle#1{\@setnumstyle\tablenum{#1}}
\def\subtablenumstyle#1{\@setnumstyle\subtablenum{#1}%
     \ifblank\subtablenumstyle\then\global\@subtblcountfalse\fi
     \ignorespaces}
\def\defnumstyle#1{\@setnumstyle\defnum{#1}}
\def\subdefnumstyle#1{\@setnumstyle\subdefnum{#1}%
     \ifblank\subdefnumstyle\then\global\@subdefcountfalse\fi
     \ignorespaces}
\def\thmnumstyle#1{\@setnumstyle\thmnum{#1}}
\def\subthmnumstyle#1{\@setnumstyle\subthmnum{#1}%
     \ifblank\subthmnumstyle\then\global\@subthmcountfalse\fi
     \ignorespaces}
\def\lemnumstyle#1{\@setnumstyle\lemnum{#1}}
\def\sublemnumstyle#1{\@setnumstyle\sublemnum{#1}%
     \ifblank\sublemnumstyle\then\global\@sublemcountfalse\fi
     \ignorespaces}

\def\heqnlabel{\newequationnum=\next
          \ifblank\subequationnumstyle\then
          \else\global\@subeqncounttrue
               \newsubequationnum=\@ne
          \fi}

\def\eqnlabel#1{%
     \if@subeqncount
          \newsubequationnum=\next
     \else\heqnlabel
     \fi
     \Label{#1}{\puteqnformat}(\puteqn{#1})%
     \ifdraft\rlap{\hskip.1in{\tt#1}}\fi}

\let\puteqn=\putlab

\def\putequation#1{\useafter\ifundefined{eqn@#1}\then
     \err@undefinedeqn{#1}\else\usename{eqn@#1}\fi}

\def\eqnseriesstyle#1{\gdef\@eqnseriesstyle{#1}}
\def\begineqnseries{\subequationnumstyle{\@eqnseriesstyle}%
     \defaultoption[]\@begineqnseries}
\def\@begineqnseries[#1]{\edef\@@eqnname{#1}}
\def\endeqnseries{\subequationnumstyle{blank}%
     \expandafter\ifnoarg\@@eqnname\then
     \else\Label\@@eqnname{\puteqnformat}%
     \fi
     \aftergroup\ignorespaces}

\def\figlabel#1{%
     \if@subfigcount
          \newsubfigurenum=\next
     \else\newfigurenum=\next
          \ifblank\subfigurenumstyle\then
          \else\global\@subfigcounttrue
               \newsubfigurenum=\@ne
          \fi
     \fi
     \Label{#1}{\putfigformat}\putfig{#1}%
   }

\let\putfig=\putlab

\def\figseriesstyle#1{\gdef\@figseriesstyle{#1}}
\def\beginfigseries{\subfigurenumstyle{\@figseriesstyle}%
     \defaultoption[]\@beginfigseries}
\def\@beginfigseries[#1]{\edef\@@figname{#1}}
\def\endfigseries{\subfigurenumstyle{blank}%
     \expandafter\ifnoarg\@@figname\then
     \else\Label\@@figname{\putfigformat}%
     \fi
     \aftergroup\ignorespaces}

\def\tbllabel#1{%
     \if@subtblcount
          \newsubtablenum=\next
     \else\newtablenum=\next
          \ifblank\subtablenumstyle\then
          \else\global\@subtblcounttrue
               \newsubtablenum=\@ne
          \fi
     \fi
     \Label{#1}{\puttblformat}\puttbl{#1}%
}

\let\puttbl=\putlab

\def\tblseriesstyle#1{\gdef\@tblseriesstyle{#1}}
\def\begintblseries{\subtablenumstyle{\@tblseriesstyle}%
     \defaultoption[]\@begintblseries}
\def\@begintblseries[#1]{\edef\@@tblname{#1}}
\def\endtblseries{\subtablenumstyle{blank}%
     \expandafter\ifnoarg\@@tblname\then
     \else\Label\@@tblname{\puttblformat}%
     \fi
     \aftergroup\ignorespaces}


\def\deflab#1{%
     \if@subdefcount
          \newsubdefnum=\next
     \else\newdefnum=\next
          \ifblank\subdefnumstyle\then
          \else\global\@subdefcounttrue
               \newsubdefnum=\@ne
          \fi
     \fi
     \Label{#1}{\putdefformat}\refdef{#1}%
}

\let\refdef=\putlab

\def\defseriesstyle#1{\gdef\@defseriesstyle{#1}}
\def\begindefseries{\subtablenumstyle{\@defseriesstyle}%
     \defaultoption[]\@begindefseries}
\def\@begindefseries[#1]{\edef\@@defname{#1}}
\def\enddefseries{\subdefnumstyle{blank}%
     \expandafter\ifnoarg\@@defname\then
     \else\Label\@@defname{\putdefformat}%
     \fi
     \aftergroup\ignorespaces}

\def\thmlab#1{%
     \if@subthmcount
          \newsubthmnum=\next
     \else\newthmnum=\next
          \ifblank\subthmnumstyle\then
          \else\global\@subthmcounttrue
               \newsubthmnum=\@ne
          \fi
     \fi
     \Label{#1}{\putthmformat}\refthm{#1}%
}

\let\refthm=\putlab

\def\thmseriesstyle#1{\gdef\@thmseriesstyle{#1}}
\def\beginthmseries{\subthmnumstyle{\@thmseriesstyle}%
     \defaultoption[]\@beginthmseries}
\def\@beginthmseries[#1]{\edef\@@thmname{#1}}
\def\endthmseries{\subthmstyle{blank}%
     \expandafter\ifnoarg\@@thmname\then
     \else\Label\@@thmname{\putthmformat}%
     \fi
     \aftergroup\ignorespaces}

\def\lemlab#1{%
     \if@sublemcount
          \newsublemnum=\next
     \else\newlemnum=\next
          \ifblank\sublemnumstyle\then
          \else\global\@sublemcounttrue
               \newsublemnum=\@ne
          \fi
     \fi
     \Label{#1}{\putlemformat}\reflem{#1}%
}

\let\reflem=\putlab

\def\lemseriesstyle#1{\gdef\@lemseriesstyle{#1}}
\def\beginlemseries{\sublemnumstyle{\@lemseriesstyle}%
     \defaultoption[]\@beginlemseries}
\def\@beginlemseries[#1]{\edef\@@lemname{#1}}
\def\endlemseries{\sublemnumstyle{blank}%
     \expandafter\ifnoarg\@@lemname\then
     \else\Label\@@lemname{\putlemformat}%
     \fi
     \aftergroup\ignorespaces}

\message{reference numbering,}

\newcount\referencenum \referencenum=0
\newcount\@@prerefcount \@@prerefcount=0
\newcount\@@thisref
\newcount\@@lastref
\newcount\@@loopref
\newcount\@@refseq
\newdimen\refnumindent
\let\@undefreflist=\empty

\def\referencenumstyle#1{\@setnumstyle\referencenum{#1}}

\def\referencestyle#1{\usename{@ref#1}}

\def\@refsequential{%
     \gdef\@refpredef##1{\global\advance\referencenum by\@ne
          \let\^^\=0\Label{##1}{\^^\{\the\referencenum}}%
          \useafter\gdef{ref@\the\referencenum}{{##1}{\undefinedlabelformat}}}%
     \gdef\@reference##1##2{%
          \ifundefinedlabel##1\then
          \else\def\^^\####1{\global\@@thisref=####1\relax}\putlab{##1}%
               \useafter\gdef{ref@\the\@@thisref}{{##1}{##2}}%
          \fi}%
     \gdef\endputreferences{%
          \loop\ifnum\@@loopref<\referencenum
                    \advance\@@loopref by\@ne
                    \expandafter\expandafter\expandafter\@printreference
                         \csname ref@\the\@@loopref\endcsname
          \repeat
          \par}}

\def\@refpreordered{%
     \gdef\@refpredef##1{\global\advance\referencenum by\@ne
          \additemR##1\to\@undefreflist}%
     \gdef\@reference##1##2{%
          \ifundefinedlabel##1\then
          \else\global\advance\@@loopref by\@ne
               {\let\^^\=0\Label{##1}{\^^\{\the\@@loopref}}}%
               \@printreference{##1}{##2}%
          \fi}
     \gdef\endputreferences{%
          \def\^^\####1{\useiflab{####1}\then
               \else\reference{####1}{\undefinedlabelformat}\fi}%
          \@undefreflist
          \par}}

\def\beginprereferences{\par
     \def\reference##1##2{\global\advance\referencenum by1\@ne
          \let\^^\=0\Label{##1}{\^^\{\the\referencenum}}%
          \useafter\gdef{ref@\the\referencenum}{{##1}{##2}}}}
\def\endprereferences{\global\@@prerefcount=\the\referencenum\par}

\def\beginputreferences{\par
     \refnumindent=\z@\@@loopref=\z@
     \loop\ifnum\@@loopref<\referencenum
               \advance\@@loopref by\@ne
               \setbox\z@=\hbox{\referencenum=\@@loopref
                    \referencenumformat\enskip}%
               \ifdim\wd\z@>\refnumindent\refnumindent=\wd\z@\fi
     \repeat
     \putreferenceformat
     \@@loopref=\z@
     \loop\ifnum\@@loopref<\@@prerefcount
               \advance\@@loopref by\@ne
               \expandafter\expandafter\expandafter\@printreference
                    \csname ref@\the\@@loopref\endcsname
     \repeat
     \let\reference=\@reference}

\def\@printreference#1#2{\ifx#2\undefinedlabelformat\err@undefinedref{#1}\fi
     \noindent\ifdraft\rlap{\hskip\hsize\hskip.1in \tt#1}\fi
     \llap{\referencenum=\@@loopref\referencenumformat\enskip}#2\par}

\def\reference#1#2{{\par\refnumindent=\z@\putreferenceformat\noindent#2\par}}

\def\putref#1{\trim#1\to\@@refarg
     \expandafter\ifnoarg\@@refarg\then
          \toks@={\relax}%
     \else\@@lastref=-\@m\def\@@refsep{}\def\@more{\@nextref}%
          \toks@={\@nextref#1,,}%
     \fi\the\toks@}
\def\@nextref#1,{\trim#1\to\@@refarg
     \expandafter\ifnoarg\@@refarg\then
          \let\@more=\relax
     \else\ifundefinedlabel\@@refarg\then
               \expandafter\@refpredef\expandafter{\@@refarg}%
          \fi
          \def\^^\##1{\global\@@thisref=##1\relax}%
          \global\@@thisref=\m@ne
          \setbox\z@=\hbox{\putlab\@@refarg}%
     \fi
     \advance\@@lastref by\@ne
     \ifnum\@@lastref=\@@thisref\advance\@@refseq by\@ne\else\@@refseq=\@ne\fi
     \ifnum\@@lastref<\z@
     \else\ifnum\@@refseq<\thr@@
               \@@refsep\def\@@refsep{,}%
               \ifnum\@@lastref>\z@
                    \advance\@@lastref by\m@ne
                    {\referencenum=\@@lastref\putrefformat}%
               \else\undefinedlabelformat
               \fi
          \else\def\@@refsep{--}%
          \fi
     \fi
     \@@lastref=\@@thisref
     \@more}

\message{streaming,}

\newif\ifstreaming

\def\streamto{\defaultoption[\jobname]\@streamto}
\def\@streamto[#1]{\global\streamingtrue
     \immediate\write\sixt@@n{--> Streaming to #1.str}%
     \newwrite\streamout\immediate\openout\streamout=#1.str }

\def\streamfrom{\defaultoption[\jobname]\@streamfrom}
\def\@streamfrom[#1]{\newread\streamin\openin\streamin=#1.str
     \ifeof\streamin
          \expandafter\err@nostream\expandafter{#1.str}%
     \else\immediate\write\sixt@@n{--> Streaming from #1.str}%
          \let\@@labeldef=\gdef
          \ifstreaming
               \edef\@elc{\endlinechar=\the\endlinechar}%
               \endlinechar=\m@ne
               \loop\read\streamin to\@@scratcha
                    \ifeof\streamin
                         \streamingfalse
                    \else\toks@=\expandafter{\@@scratcha}%
                         \immediate\write\streamout{\the\toks@}%
                    \fi
                    \ifstreaming
               \repeat
               \@elc
               \input #1.str
               \streamingtrue
          \else\input #1.str
          \fi
          \let\@@labeldef=\xdef
     \fi}

\def\streamcheck{\ifstreaming
     \immediate\write\streamout{\pagenum=\the\pagenum}%
     \immediate\write\streamout{\footnotenum=\the\footnotenum}%
     \immediate\write\streamout{\referencenum=\the\referencenum}%
     \immediate\write\streamout{\chapternum=\the\chapternum}%
     \immediate\write\streamout{\sectionnum=\the\sectionnum}%
     \immediate\write\streamout{\subsectionnum=\the\subsectionnum}%
     \immediate\write\streamout{\equationnum=\the\equationnum}%
     \immediate\write\streamout{\subequationnum=\the\subequationnum}%
     \immediate\write\streamout{\figurenum=\the\figurenum}%
     \immediate\write\streamout{\subfigurenum=\the\subfigurenum}%
     \immediate\write\streamout{\tablenum=\the\tablenum}%
     \immediate\write\streamout{\subtablenum=\the\subtablenum}%
     \immediate\closeout\streamout
     \fi}


\def\err@badtypesize{%
     \errhelp={The limited availability of certain fonts requires^^J%
          that the base type size be 10pt, 12pt, or 14pt.^^J}%
     \errmessage{--> Illegal base type size}}

\def\err@badsizechange{\immediate\write\sixt@@n
     {--> Size change not allowed in math mode, ignored}}

\def\err@sizetoolarge#1{\immediate\write\sixt@@n
     {--> \noexpand#1 too big, substituting HUGE}}

\def\err@sizenotavailable#1{\immediate\write\sixt@@n
     {--> Size not available, \noexpand#1 ignored}}

\def\err@fontnotavailable#1{\immediate\write\sixt@@n
     {--> Font not available, \noexpand#1 ignored}}

\def\err@sltoit{\immediate\write\sixt@@n
     {--> Style \noexpand\sl not available, substituting \noexpand\it}%
     \it}

\def\err@bfstobf{\immediate\write\sixt@@n
     {--> Style \noexpand\bfs not available, substituting \noexpand\bf}%
     \bf}

\def\err@badgroup#1#2{%
     \errhelp={The block you have just tried to close was not the one^^J%
          most recently opened.^^J}%
     \errmessage{--> \noexpand\End{#1} doesn't match \noexpand\begin{#2}}}

\def\err@badcountervalue#1{\immediate\write\sixt@@n
     {--> Counter (#1) out of bounds}}

\def\err@extrafootnotemark{\immediate\write\sixt@@n
     {--> \noexpand\footnotemark command
          has no corresponding \noexpand\footnotetext}}

\def\err@extrafootnotetext{%
     \errhelp{You have given a \noexpand\footnotetext command without first
          specifying^^Ja \noexpand\footnotemark.^^J}%
     \errmessage{--> \noexpand\footnotetext command has no corresponding
          \noexpand\footnotemark}}

\def\err@labelredef#1{\immediate\write\sixt@@n
     {--> Label "#1" redefined}}

\def\err@badlabelmatch#1{\immediate\write\sixt@@n
     {--> Definition of label "#1" doesn't match value in \jobname.lab}}

\def\err@needlabel#1{\immediate\write\sixt@@n
     {--> Label "#1" cited before its definition}}

\def\err@undefinedlabel#1{\immediate\write\sixt@@n
     {--> Label "#1" cited but never defined}}

\def\err@undefinedeqn#1{\immediate\write\sixt@@n
     {--> Equation "#1" not defined}}

\def\err@undefinedref#1{\immediate\write\sixt@@n
     {--> Reference "#1" not defined}}

\def\err@nostream#1{%
     \errhelp={You have tried to input a stream file that doesn't exist.^^J}%
     \errmessage{--> Stream file #1 not found}}

\message{jyTeX initialization}

\everyjob{\immediate\write16{--> jyTeX version \fmtversion}%
     \edef\@@jobname{\jobname}%
     \edef\jobname{\@@jobname}%
     \settime
     \openin0=\jobname.lab
     \ifeof0
     \else\closein0
          \immediate\write16{--> Getting labels from file \jobname.lab}%
          \input\jobname.lab
     \fi}


%
     \^^\{\splittopskip}%
     \^^\{\maxdepth}%
     \^^\{\skip\topins}%
     \^^\{\skip\footins}%
     \^^\{\headskip}%
     \^^\{\footskip}}

\def\scalingskipslist{%
     \^^\{\p@renwd}%
     \^^\{\delimitershortfall}%
     \^^\{\nulldelimiterspace}%
     \^^\{\scriptspace}%
     \^^\{\jot}%
     \^^\{\normalbaselineskip}%
     \^^\{\normallineskip}%
     \^^\{\normallineskiplimit}%
     \^^\{\baselineskip}%
     \^^\{\lineskip}%
     \^^\{\lineskiplimit}%
     \^^\{\bigskipamount}%
     \^^\{\medskipamount}%
     \^^\{\smallskipamount}%
     \^^\{\parskip}%
     \^^\{\parindent}%
     \^^\{\abovedisplayskip}%
     \^^\{\belowdisplayskip}%
     \^^\{\abovedisplayshortskip}%
     \^^\{\belowdisplayshortskip}%
     \^^\{\abovechapterskip}%
     \^^\{\belowchapterskip}%
     \^^\{\abovesectionskip}%
     \^^\{\belowsectionskip}%
     \^^\{\abovesubsectionskip}%
     \^^\{\belowsubsectionskip}}


\def\twoupsetup{
     \topmargin=.75in
     \leftmargin=.5in
     \vsize=6.9in
     \hsize=4.75in
     \fullhsize=10in
     \let\draft=\relax}


\chapterstyle{left}                              
\chapternumstyle{blank}                          
\def\chapterbreak{\newpage}                      
\abovechapterskip=0pt                            
\belowchapterskip=1.5\baselineskip               
     plus.38\baselineskip minus.38\baselineskip
\def\chapternumformat{\numstyle\chapternum.}     

\sectionstyle{left}                              
\sectionnumstyle{blank}                          
\def\sectionbreak{\vskip0pt plus4\baselineskip\penalty-100
     \vskip0pt plus-4\baselineskip}              
\abovesectionskip=1.5\baselineskip               
     plus.38\baselineskip minus.38\baselineskip
\belowsectionskip=\the\baselineskip              
     plus.25\baselineskip minus.25\baselineskip
\def\sectionnumformat{
     \ifblank\chapternumstyle\then\else\numstyle\chapternum.\fi
     \numstyle\sectionnum.}

\subsectionstyle{left}                           
\subsectionnumstyle{blank}                       
\def\subsectionbreak{\vskip0pt plus4\baselineskip\penalty-100
     \vskip0pt plus-4\baselineskip}              
\abovesubsectionskip=\the\baselineskip           
     plus.25\baselineskip minus.25\baselineskip
\belowsubsectionskip=.75\baselineskip            
     plus.19\baselineskip minus.19\baselineskip
\def\subsectionnumformat{
     \ifblank\chapternumstyle\then\else\numstyle\chapternum.\fi
     \ifblank\sectionnumstyle\then\else\numstyle\sectionnum.\fi
     \numstyle\subsectionnum.}


\def\undefinedlabelformat{$\bullet$}             


\equationnumstyle{arabic}                        
\subequationnumstyle{blank}                      
\figurenumstyle{arabic}                          
\subfigurenumstyle{blank}                        
\tablenumstyle{arabic}                           
\subtablenumstyle{blank}                         
\defnumstyle{arabic}                             
\subdefnumstyle{blank}                           
\thmnumstyle{arabic}                             
\subthmnumstyle{blank}                           
\lemnumstyle{arabic}                             
\sublemnumstyle{blank}                           

\eqnseriesstyle{alphabetic}                      
\figseriesstyle{alphabetic}                      
\tblseriesstyle{alphabetic}                      
\defseriesstyle{alphabetic}                      
\thmseriesstyle{alphabetic}                      
\lemseriesstyle{alphabetic}                      

\def\puteqnformat{\hbox{
     \ifblank\chapternumstyle\then\else\numstyle\chapternum.\fi
     \ifblank\sectionnumstyle\then\else\numstyle\sectionnum.\fi
     \ifblank\subsectionnumstyle\then\else\numstyle\subsectionnum.\fi
     \numstyle\equationnum
     \numstyle\subequationnum}}
\def\putfigformat{\hbox{
     \ifblank\chapternumstyle\then\else\numstyle\chapternum.\fi
     \ifblank\sectionnumstyle\then\else\numstyle\sectionnum.\fi
     \ifblank\subsectionnumstyle\then\else\numstyle\subsectionnum.\fi
     \numstyle\figurenum
     \numstyle\subfigurenum}}
\def\puttblformat{\hbox{
     \ifblank\chapternumstyle\then\else\numstyle\chapternum.\fi
     \ifblank\sectionnumstyle\then\else\numstyle\sectionnum.\fi
     \ifblank\subsectionnumstyle\then\else\numstyle\subsectionnum.\fi
     \numstyle\tablenum
     \numstyle\subtablenum}}
\def\putdefformat{\hbox{
     \ifblank\chapternumstyle\then\else\numstyle\chapternum.\fi
     \ifblank\sectionnumstyle\then\else\numstyle\sectionnum.\fi
     \ifblank\subsectionnumstyle\then\else\numstyle\subsectionnum.\fi
     \numstyle\defnum
     \numstyle\subdefnum}}
\def\putthmformat{\hbox{
     \ifblank\chapternumstyle\then\else\numstyle\chapternum.\fi
     \ifblank\sectionnumstyle\then\else\numstyle\sectionnum.\fi
     \ifblank\subsectionnumstyle\then\else\numstyle\subsectionnum.\fi
     \numstyle\thmnum
     \numstyle\subthmnum}}
\def\putlemformat{\hbox{
     \ifblank\chapternumstyle\then\else\numstyle\chapternum.\fi
     \ifblank\sectionnumstyle\then\else\numstyle\sectionnum.\fi
     \ifblank\subsectionnumstyle\then\else\numstyle\subsectionnum.\fi
     \numstyle\lemnum
     \numstyle\sublemnum}}


\referencestyle{sequential}                      
\referencenumstyle{arabic}                       
\def\putrefformat{\numstyle\referencenum}        
\def\referencenumformat{\numstyle\referencenum.} 
\def\putreferenceformat{
     \everypar={\hangindent=1em \hangafter=1 }%
     \def\\{\hfil\break\null\hskip-1em \ignorespaces}%
     \leftskip=\refnumindent\parindent=0pt \interlinepenalty=1000 }


\def\fmtversion{2.6M (June 1992)}

\catcode`\@=12

\def\ref#1{(\puteqn{#1})}
\def\label#1{\eqno\eqnlabel{#1}}
\font\bigboldfont=cmbx10 scaled \magstep2
\def\displayhead#1{{\bigboldfont \leftline{#1}}
\vskip-10pt
\line{\hrulefill}}
\def\section#1{\ifblank\sectionnumstyle\then
          \else\newsectionnum=\next \fi
\displayhead{\ifblank\sectionnumstyle\then\else\sectionnumformat\ \fi#1}
     }
\def\appendix#1{\ifblank\sectionnumstyle\then
          \else\newsectionnum=\next \fi
\displayhead{Appendix
    \ifblank\sectionnumstyle\then\else\sectionnumformat\ \fi#1}
     }


%
\input psfig.sty
\chapternumstyle{blank}                           
\sectionnumstyle{arabic}                          
\def\cite#1{$\lbrack#1\rbrack$}
\def\bibitem#1{\parindent=9mm\item{\hbox to 7 mm{\cite{#1}\hfill}}}
\def\BTWa{1}
\def\BTWb{2}
\def\BCT{3}
\def\BakS{4}
\def\FBS{5}
\def\GrKa{6}
\def\DMS{7}
\def\DrSchwA{8}
\def\LPVZ{9}
\def\DrSchF{10}
\def\CDrSchE{11}
\def\BakPac{12}
\def\CDrSchw{13}
\def\Henley{14}
\def\ADHR{15}
\def\CDrSchwA{16}
\def\BJW{17}
\def\Grassb{18}
\def\FYWu{19}
\def\GJS{20}
\def\Drossel{21}
\def\PaTri{22}
\def\BoDe{23}
\def\SchueHen{24}
\def\BuHe{25}
\def\CDrSchP{26}
\def\sn{\smallskip\noindent}
\def\mn{\medskip\noindent}
\def\bn{\bigskip\noindent}
\def\min{{\rm min}}
\def\max{{\rm max}}

\def\Cor{C}
\def\Hs{H_{\rm s}}
\def\Eex{\Lambda_{\rm ex}}
\def\nuT{\nu_T}
\def\Es{0}
\def\Esl{\mathop{\hbox{$\Es$}}\limits}
\def\Ts{1}
\def\ffm{ffm}
\def\oh{{\textstyle {1 \over 2}}}
\def\Epr{\oh \left(\id - \sigma^z\right)}
\def\Tpr{\oh \left(\id + \sigma^z\right)}
\def\sip{\sigma^{+}}
\def\simi{\sigma^{-}}
\def\Textfrac#1#2{#1/#2}
\def\textfrac#1#2{\Textfrac{#1}{\left(#2\right)}}
\def\figindents{\leftskip=4.5 true pc \rightskip=4 true pc}

\def\sn{\smallskip\noindent}

\def\abs#1{\vert #1 \vert}

\def\state#1{\,\hbox{$\mid \! #1 \rangle$}\, }
\def\rrangle{\rangle \hskip -1pt \rangle}

\def\astate#1{\,\hbox{$\langle #1 \! \mid$}\, }
\def\pstate#1{\Vert #1 \rrangle\, }

\def\tstate#1{\state{\widetilde{#1}}}
\def\atstate#1{\astate{\widetilde{#1}}}

\def\GS{\state{G}}
\def\Order#1{{\cal O}(#1)}
\font\ninerm=cmr8
\font\nineit=cmmi8
\font\lasy=lasy10
\chardef\SymbolAchar="32
\chardef\SymbolBchar="33
\def\SymbolA{\rlap{\kern0.8mm\raise0.3mm\hbox{$
           \cdot$}}\hbox{\lasy\SymbolAchar}}
\def\SymbolB{\rlap{\kern0.9mm\raise0.3mm\hbox{$
           \cdot$}}\hbox{\lasy\SymbolBchar}}
\font\linew=linew10
\font\circle=lcircle10
\chardef\ArrowNum="1B
\chardef\RArrowNum="2D
\chardef\QcircANum="26
\chardef\QcircBNum="25
\def\Arrow{\linew \ArrowNum}
\def\RArrow{\linew \RArrowNum}
\def\QcircA{\circle \QcircANum}
\def\QcircB{\circle \QcircBNum}
\def\verline#1#2#3{\rlap{\kern#1mm\raise#2mm
                   \hbox{\vrule height #3mm depth 0 pt}}}
\def\horline#1#2#3{\rlap{\kern#1mm\raise#2mm
                   \vbox{\hrule width #3mm depth 0 pt}}}
\def\Verline#1#2#3{\rlap{\kern#1mm\raise#2mm
                   \hbox{\vrule height #3mm width 0.7pt depth 0 pt}}}
\def\Horline#1#2#3{\rlap{\kern#1mm\raise#2mm
                   \vbox{\hrule height 0.7pt width #3mm depth 0 pt}}}
\def\putbox#1#2#3{\setbox117=\hbox{#3}
                  \dimen121=#1mm
                  \dimen122=#2mm
                  \dimen123=\wd117
                  \dimen124=\ht117
                  \divide\dimen123 by -2
                  \divide\dimen124 by -2
                  \advance\dimen121 by \dimen123
                  \advance\dimen122 by \dimen124
                  \rlap{\kern\dimen121\raise\dimen122\hbox{#3}}}
\def\leftputbox#1#2#3{\setbox117=\hbox{#3}
                  \dimen122=#2mm
                  \dimen124=\ht117
                  \divide\dimen124 by -2
                  \advance\dimen122 by \dimen124
                  \rlap{\kern#1mm\raise\dimen122\hbox{#3}}}
\def\topputbox#1#2#3{\setbox117=\hbox{#3}
                  \dimen121=#1mm
                  \dimen123=\wd117
                  \divide\dimen123 by -2
                  \advance\dimen121 by \dimen123
                  \rlap{\kern\dimen121\raise#2mm\hbox{#3}}}
\def\point#1#2{
  \putbox{#1}{#2}{$\bullet$}}

\setbox22 = \hbox{1}
\def\id{\rlap{1}\rlap{\kern 1pt \vbox{\hrule width 4pt depth 0 pt}}
        \rlap{\kern 3.5 pt \hbox{\vrule height \ht22 depth 0 pt}}
            \hskip\wd22}
\font\amsmath=msbm10

\def\Zed{\hbox{\amsmath Z}}

%
%
\font\large=cmbx10 scaled \magstep3
\font\bigf=cmr10 scaled \magstep2
\pageno=0
\def\folio{
\ifnum\pageno<1 \footline{\hfil} \else\number\pageno \fi}
\rightline{cond-mat/9511132}
\vskip 3.0truecm
\centerline{\large Critical Properties of the}
\vskip 0.5truecm
\centerline{\large One-Dimensional Forest-Fire Model}
\vskip 1.5truecm
\centerline{\bigf A.\ Honecker and I.\ Peschel}
\bigskip\medskip
\centerline{\it Fachbereich Physik, Freie Universit\"at Berlin,}
\centerline{\it Arnimallee 14, D--14195 Berlin, Germany}
\vskip 1.9truecm
\centerline{\bf Abstract}
\vskip 0.2truecm
\noindent
The one-dimensional
forest-fire model including lightnings is studied numerically
and analytically. For the tree correlation function, a new
correlation length with critical exponent $\nu \approx 5/6$
is found by simulations. A Hamiltonian formulation is introduced
which enables one to study the stationary state close to
the critical point using quantum-mechanical perturbation
theory. With this formulation also the structure of the low-lying
relaxation spectrum and the critical behaviour of the smallest
complex gap are investigated numerically. Finally, it is
shown that critical correlation functions can be
obtained from a simplified model involving only the total
number of trees although such simplified models are unable
to reproduce the correct off-critical behaviour.
\vfill
\leftline{\hbox to 5 true cm{\hrulefill}}
\leftline{e-mail:}
\leftline{\quad honecker@omega.physik.fu-berlin.de}
\leftline{\quad peschel@aster.physik.fu-berlin.de}
\eject
\section{Introduction}
\mn
Power laws and scaling behaviour, familiar from equilibrium
critical phenomena, can also be found in non-equilibrium systems.
If in such a case no fine-tuning of a parameter is necessary,
the situation has been described as self-organized criticality
\cite{\BTWa,\BTWb}. To illustrate the phenomenon, simple models
for sandpiles \cite{\BTWa,\BTWb}, forest fires \cite{\BCT} or evolution
\cite{\BakS,\FBS} have been proposed. Their common feature are
avalanche-type processes in the dynamics. In the case of the
forest-fire models (ffm), the originally proposed version did
not show proper scaling behaviour \cite{\GrKa,\DMS}, and it was
necessary to introduce lightning strokes in order to find
it \cite{\DrSchwA}. Even then, the spatial power laws are valid
only up to a typical length depending on the lightning rate.
Therefore the ffm is in general not at a critical point, but
only close to it. This is also the picture obtained from a
renormalization treatment \cite{\LPVZ}. Nevertheless, the ffms are
interesting systems (for a brief review see \cite{\DrSchF} and for
a recent account \cite{\CDrSchE}), and a detailed understanding of
their properties, in particular in comparison with usual critical
systems, is desirable.
This should be simplest in the one-dimensional case.
\mn
In one dimension, a few exact results have been obtained \cite{\BakPac,
\CDrSchw}. In partiuclar, the stationary distribution of tree clusters
with size $s$ is given by \cite{\CDrSchw}
$$n(s) = {(1 - \rho) \over (s+1) (s+2)}
\label{EQns}$$
where $\rho$ is the density. This expression is valid for $s \ll \xi_c$
and gives $n(s) \sim s^{-2}$ for large $s$. The length $\xi_c$
depends on the rates $p$ (for tree growth) and $f$ (for lightning
strokes) as $\xi_c \sim p/f$, up to logarithmic corrections.
Thus, the corresponding exponent is $\nu = 1$. The time correlation
function of burning trees could also be obtained. These results were
checked by numerical calculations and a certain overall picture
has emerged. However, the precise nature of the stationary state
or the form of the relaxation spectrum are still to be determined.
\mn
To make some progress in this direction, we first studied the
stationary correlations in more detail. In particular, we
looked at the simple tree correlation function and found an
unexpected result, namely a different correlation length which
diverges with an exponent $\nuT \approx \Textfrac{5}{6}$. This is in
contrast to results in two dimensions \cite{\Henley} and
indicates a rather complicated structure of the stationary state.
To obtain this state explicitly, one has to treat the master
equation. For this we used a quantum-mechanical formulation which
has proved to be quite useful in other problems (see for example
\cite{\ADHR}). If one
assumes instant burning of tree clusters, one then obtains a spin
one-half quantum chain with both single and cluster-flip
processes. A number of eigenstates can be found exactly, but
in general one has to resort to numerics. We calculated the
low-lying spectrum for systems up to $L=20$ sites, obtaining
various branches in the complex plane and a gap which closes
algebraically as the rate $f$ goes to zero. We also considered
the limit $f \to 0$ analytically and found that to order
${\cal O}(f)$ the ground state lies in a subspace of $L+1$
totally symmetric configurations. Working only in this
subspace, one obtains a simple model where the system evolves
in cycles \cite{\CDrSchwA}. Then the stationary state can be
found explicitly and gives the cluster
distribution \ref{EQns} but it contains no correlations.
Also the low-lying spectrum differs from the exact one.
A similar observation for models of evolution has already
been made in \cite{\BJW} where it was concluded that
the behaviour of only one quantity is not a sufficient
criterion for criticality. To pursue this aspect further,
we also investigated to what extent the cluster-size
distribution of the full model for $f > 0$ can be
recovered in the symmetric subspace.
\bn
\section{Simulations of correlation functions}
\mn
This section presents results of simulations of two correlation
functions. First, the two-point function of trees is studied
and it is shown that the exponent of the corresponding
correlation length is not an integer. By contrast, the
correlation function inside a single cluster shows a
critical behaviour which is consistent with
the critical exponent of the cluster-size distribution
$n(s)$ eq.\ \ref{EQns}.
\mn
The \ffm{} introduced in \cite{\BCT} is defined
on a cubic lattice in $d$ dimensions. Any site can have three
states: It can be empty, or it can either be occupied by a
green tree or a by a burning tree. The dynamics of the model
was specified in \cite{\BCT} by the following parallel update
rules from one timestep to the next:
a) A burning tree becomes an empty site.
b) A green tree becomes a burninng tree if at least one of its
   nearest neighbours is burning.
c) At an empty site a green tree grows with probability $p$.
In order to obtain proper critical behaviour, the following
rule was added in \cite{\DrSchwA}:
d) Every green tree can spontaneously be struck by lightning
   with probability $f$ and thus become a burning tree even if no
   neighbour is burning.
\mn
Quantities of interest in this model are in particular the
forest clusters, i.e.\ maximal connected sets of green trees.
In one dimension a forest cluster is a string of green trees
bounded by empty sites and its size $s$ is just the number of trees.
\mn
The critical point of the above \ffm{}, which has
been studied as an example of self-organized criticality,
arises at $p \to 0$ and $f/p \to 0$. Keeping $f/p$ fixed and
taking $p \to 0$ has the effect that once a cluster is struck
by lightning it burns down before anything else happens.
After redefining the time scale, the problem reduces to a
two-state model where only empty sites ($E$) and trees ($T$)
occur. The dynamics of this effective model is specified by
the following parallel update rules:
\item{1)} At an empty site a tree grows with probability $p$.
\item{2)} A forest cluster of size $s$ is struck by lightning
          with rate $f s$ and all trees inside it become empty
          sites during one timestep.
\par\noindent
Simulations have been carried out almost always with this reduced
model (see e.g.\ \cite{\Grassb,\Henley}). Below we will always
think of this reduced version if we refer to `the forest-fire model'.
We will restrict our attention to the one-dimensional version
with periodic boundary conditions.
The size of the lattice will be denoted by $L$.
\mn
First we have investigated two-point functions of trees
$\langle T_x T_{x+y}\rangle$ in one dimension.
To this end we have performed simulations for $0.0025 \le p \le 0.01$
in the range $2 \, 10^{-5} \le f/p \le 2 \, 10^{-2}$. The sizes of lattices
ranged between $L=4000$ for large $f$ and $L=500000$ for small $f$.
One random initial condition with density $\rho = {1 \over 2}$ was chosen.
Then $t_0$ iterations were performed in order to equilibrate the system
(usually $t_0 = 3000$, only for very small $f$ one needs a few
more iterations). After that $\langle T_x T_{x+y}\rangle$ was determined
every 100 timesteps from $t_0$ to $t_0 + 1900$ by averaging over all $x$.
This determination of $\langle T_x T_{x+y}\rangle$ is
much more time consuming than the simulation itself. The procedure
was repeated 10 times with the same initial conditions, amounting
to a total of $200 L$ measurements for $\langle T_x T_{x+y}\rangle$.
\mn
The resulting data can nicely be fitted by
$$\Cor(y) := \langle T_x T_{x+y}\rangle - \langle T_x \rangle^2
        = a e^{-\Textfrac{y}{\xi}}
\label{TPform}$$
for $1 \le y$. (Note that $\langle T_x \rangle = \rho$ and $T_x^2 = T_x$.)
The parameters $a$ and $\xi$ were estimated by taking the logarithm of
the r.h.s.\ of \ref{TPform} and then performing a linear regression for
$1 \le y \le y_{\rm max}$. $y_{\rm max}$ was chosen such that
statistical errors can be neglected for $y < y_{\rm max}$. With the
statistics of our simulations $y_{\rm max}$ was always a few correlation
lengths.
\mn
\centerline{\psfig{figure=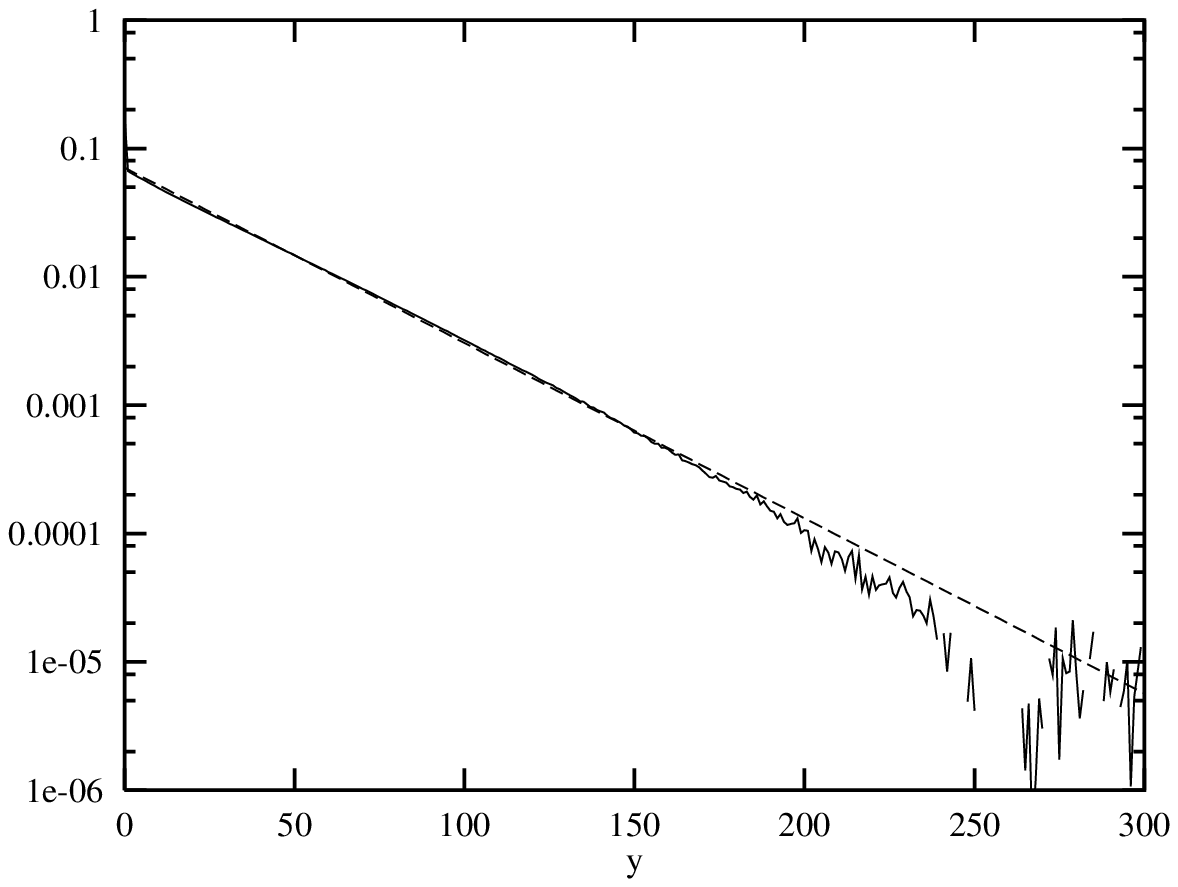}}
\sn
{\par\noindent\figindents
{\bf Fig.\ 1:}
The correlation function $\Cor(y)$ for $L=100000$, $p = 1/200$,
$f/p = 1/100$ (full line). The dashed line is the exponential
form \ref{TPform} with $a=0.0707$, $\xi = 31.81$.
\par\noindent}
\mn
Fig.\ 1 shows the result of a detailed simulation
where the average was
taken every 100 timesteps over the larger time interval from
$t_0 = 3000$ to $t_0 + 39900$. The agreement of the simulation
with the exponential form is very good for a few
correlation lengths. For $y \approx 200$ statistical errors start
to become so large that the difference on the l.h.s.\ of \ref{TPform}
cannot be determined sufficiently accurately any more.
For smaller values of $y$ there are small wiggles over
longer ranges of the distance $y$ which are typical for
the simulations and seem to be caused by statistical fluctuations.
Apart from that, no substantial
corrections to the form \ref{TPform} are visible.
\mn
Fig.\ 2 shows the results of the correlation length $\xi$ obtained
from simulations as a function of $f/p$. There are some residual
statistical as well as systematic errors due to a small $p$-dependence
which are, however, not larger than the symbols. One can see
that the values for $\xi$ (and similarly for $a$)
are in good agreement with the form
$$\xi \sim \left({f \over p}\right)^{-\nuT} \, , \qquad
a \sim \left({f \over p}\right)^{\mu} \, .
\label{TPcrit}$$
Performing linear regression fits on a doubly logarithmic scale
one finds
$$\nuT = 0.8336 \pm 0.0036 \, , \qquad
\mu = 0.1031 \pm 0.0022 \, .
\label{TPexp}$$
Surprisingly $\nuT$ agrees very well with the gap index
$\nu = \Textfrac{5}{6}$ for the three-state Potts model \cite{\FYWu}.
The most probable rational value for the other exponent is
$\mu = \Textfrac{1}{10}$.
\mn
\centerline{\psfig{figure=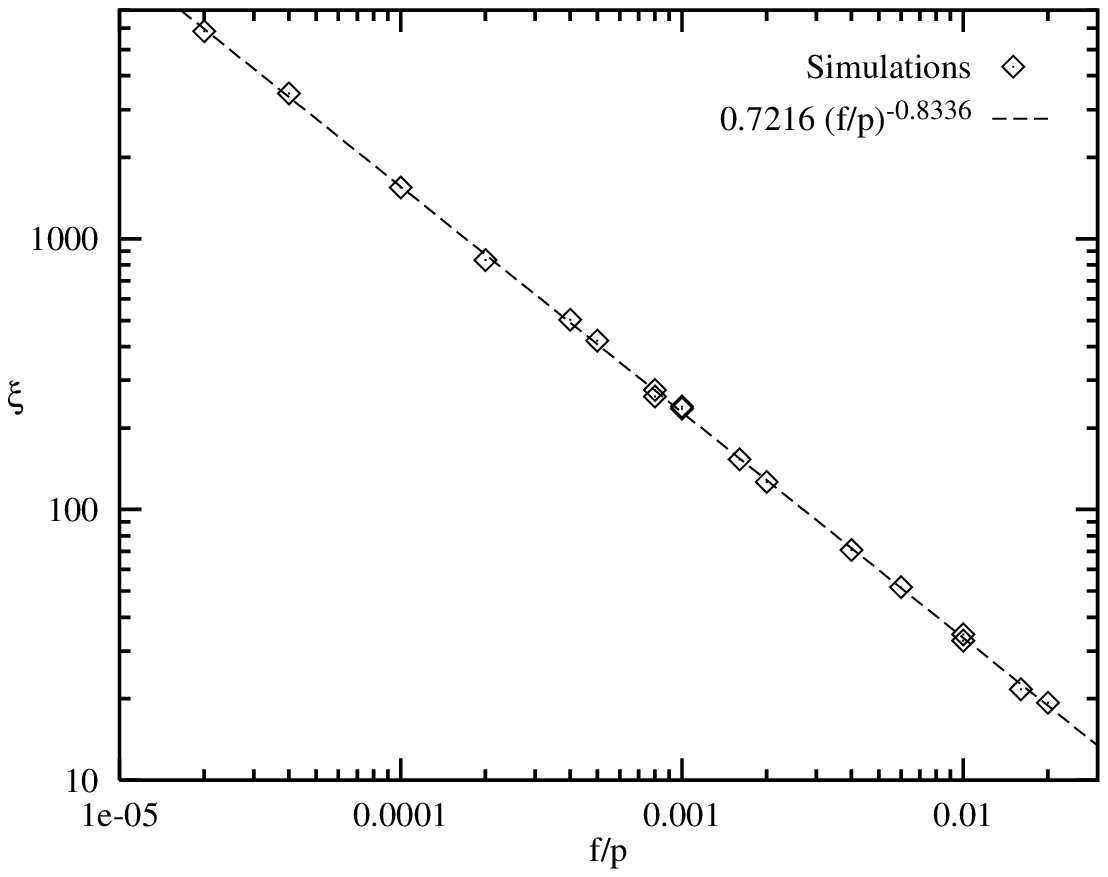}}
\sn
{\par\noindent\figindents
{\bf Fig.\ 2:} The correlation length $\xi$ as a function of $f/p$
for various values of $f$, $p$ and $L$.
\par\noindent}
\mn
It was argued in \cite{\Henley} for two dimensions that the
critical exponents are independent of the particular correlation
function considered. In order to investigate this question in
one dimension we also looked at the correlation function
$\langle T_x T_{x+y}\rangle_c$ describing the probability to find two
trees at positions $x$ and $x+y$ {\it inside the same cluster}.
In $d=1$ one has
$$\langle T_x T_{x+y}\rangle_c =
\langle T_x T_{x+1} \ldots T_{x+y}\rangle =: K(y) \, .
\label{CONdef}$$
This quantity can easily be simulated as before. Its
determination is actually much less time consuming.
\mn
The correlation function $K(y)$ can be fitted with a single
exponential function for $y$ sufficiently large, but for smaller
$y$ an exponential function is not a good approximation. Thus,
we fit $K(y)$ by
$$K(y) =
  \sum_{r=1}^{\infty} a^{(r)} e^{-\Textfrac{y}{\xi_c^{(r)}}}
\label{CONform}$$
where the $\xi_c^{(r)}$ decrease with $r$. The first few terms
in the sum \ref{CONform} already give excellent approximations.
In an interval $y_\min \le y \le y_\max$ all terms with $r > 1$
can be neglected and this can be used
to estimate the largest decay length $\xi_c = \xi_c^{(1)}$ with a linear
regression for the logarithm of the correlation function $K(y)$.
Typically one finds $y_\min \approx \xi_c$, $y_\max \ge 5 \xi_c$.
\mn
Note that $K(y)$ is related to the cluster-size distribution $n(s)$ via
$$K(y) = \sum_{s > y} (s-y) n(s) \, .
\label{CONns}$$
This can be inverted to give
$$n(s) = K(s-1) - 2 K(s) + K(s+1) \, .
\label{CONnsK}$$
This implies that if $K(s)$ decays exponentially in some
region, also $n(s)$ must decay exponentially in roughly the
same region. However, the lattice Laplacian in \ref{CONnsK}
increases the weight of the next to leading terms in
\ref{CONform}. Therefore, the purely exponential decay is
expected to set in at larger $s$ for $n(s)$ than for $K(s)$.
The value $s_0$ where a single exponential function starts
to describe $n(s)$ can be estimated as follows. One requires that
the form $n(s) = \Textfrac{(1-\rho)}{s^2}$ can be matched up to the
linear term at $s_0$ with a single exponential function for
$K(s)$, i.e.\ $K(s) = a e^{-\Textfrac{s}{\xi_c}}$. According
to \ref{CONnsK} one must have
$n(s_0) = {\partial^2 \over \partial s^2} K(s)\vert_{s=s_0}$
and ${\partial \over \partial s} n(s) \vert_{s=s_0} =
{\partial^3 \over \partial s^3} K(s)\vert_{s=s_0}$.
This leads to the relation $a \approx (1 - \rho)$
(which is indeed well verified in simulations) for
the normalization constants and $s_0 \approx \xi_c$
for the desired crossover point.
\mn
Fig.\ 3 shows the correlation function $K(s)$ and the
normalized cluster-size distribution $n(s)$. The simulation
was performed in the same manner as in Fig.\ 1. $n(s)$
was determined by counting all clusters of size $s$ in
the system during each of the $40000$ timesteps
after $t_0 = 3000$. In this time-interval, estimates for
$K(y)$ were obtained every hundreth timestep
as a spatial average over the entire system. The huge amount of
data in particular for $n(s)$ is needed in order to
obtain small statistical fluctuations without too much
extra smoothing. One observes that
both expectation values decay exponentially over
a large range of $s$. The correlation lengths obtained
by fits in this region agree very well with each
other as it should be according to \ref{CONnsK}.
The actual value is $\xi_c \approx 28$.
Furthermore, the crossover of the cluster-size
distribution from the form \ref{EQns}
to the exponential decay does indeed occur in the vicinity
of one correlation length, the exponential decay
being well visible beyond $s = 10 \xi_c$.
It seems that $K(y)$ starts to decay faster than
exponentially around $s=400$ in Fig.\ 3.
However, comparison with simulations with a
smaller amount of samples indicates that this effect
is a residual statistical error. Thus, the exponential
decay of $n(s)$ and $K(s)$ could well extend indefinitely,
the probability of clusters larger than 450 trees simply
being so small that they did not appear in the
present simulation.
\mn
Although the simulation in Fig.\ 3 was performed for
a fairly large value of $f / p$, it does reflect
the typical situation also closer to the critical point
$f / p = 0$. The form \ref {EQns} for $n(s)$
is valid approximately up to one correlation length $\xi_c$,
while the exponential decay can be verified for at least
a few $\xi_c$. For smaller $f / p$ than in Fig.\ 3
there is a small difference: There is a region
where the form \ref{EQns} for $n(s)$ lies below the
actual exponential decrease. This corresponds to
the `bump' observed in a logarithmic plot of the cluster
size distribution \cite{\CDrSchw}.
\mn
\centerline{\psfig{figure=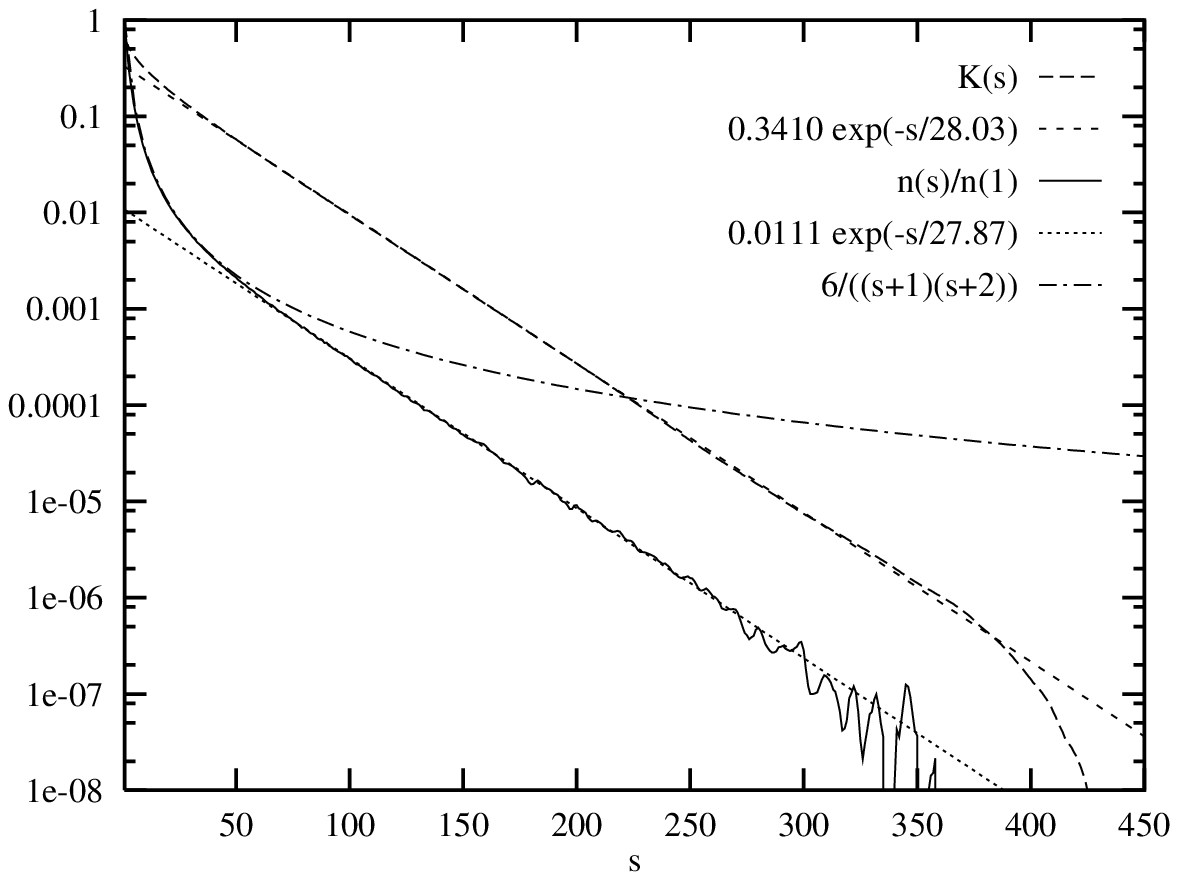}}
\sn
{\par\noindent\figindents
{\bf Fig.\ 3:} The correlation function $K(s)$ and
the cluster-size distribution $n(s)$ for $L=100000$,
$p = 1/200$, $f/p = 1/100$.
\par\noindent}
\mn
There is a striking similarity of $n(s)$ in Fig.\ 3 with
the avalanche distribution shown in Fig.\ 6b of \cite{\GJS}
for a different two-dimensional forest-fire model. This
similarity might indicate that one could expect similar
results for the two-dimensional version of the forest fire
model considered in this paper, but it may also just
illustrate the fact that cluster size distributions of
very different models can qualitatively resemble each other.
\mn
We have also performed simulations to study the $f/p$-dependence
of $K(y)$. They were done for $4 \; 10^{-4} \le f/p \le 2\; 10^{-2}$
and $15000 \le L \le 500000$. One finds that the resulting data is
consistent with
$$\xi_c \; \ln(\xi_c) = (0.95 \pm 0.06) \, {p \over f} \, .
\label{CONexp}$$
The functional form of the relation \ref{CONexp} was already
given in Ref.\ \cite{\CDrSchw} although the derivation
has to be considered with some care (see section 5 below).
\mn
This result is to be contrasted with the one for the two-point
correlation function $C(y)$. The two-point function
is significantly more accurately descibed by a single
exponential function (compare Figs.\ 1 and 3). Even more,
the critical exponents are clearly different -- see eq.\ \ref{TPexp}.
Thus, in one dimension there seems to be no direct relation
between the two different correlation functions.
Note that the relative errors in Ref.\ \cite{\Henley} for the critical
exponents in two dimensions are much larger than the ones of our
simulations. Therefore it is conceivable that a more precise
study could reveal different exponents in two dimensions as well.
\mn
Finally, we would like to mention than an exponential decay of $n(s)$
as found above, can already be obtained from a mean-field
approximation. Complete factorization of the correlator defining $n(s)$
yields $n(s) = (1-\rho)^2 \rho^s$, i.e.\ $n(s)$ varies
exponentially. In the mean-field approximation for the original version
of the \ffm{} with burning trees one finds the critical behaviour
$1 - \rho \sim f/p$. This implies $\xi_c = -\Textfrac{1}{\ln(\rho)}
\sim p/f$, i.e.\ the critical exponent is $\nu = 1$ in mean-field
approximation. This agrees with \ref{CONexp} up to logarithmic corrections
but one should note that the critical behaviour of $\rho$ is not
right. The correct result is $\textfrac{\rho}{1-\rho} \sim \ln(p/f)$
\cite{\CDrSchw} and the corresponding exponent differs by one from the
mean-field result.
\bn
\section{Hamiltonian formulation of the model}
\mn
In this section we present a quantum-mechanical formulation
of the \ffm{} and use it to discuss correlations
at the critical point.
\mn
In order to set up notations, let us briefly recall the Hamiltonian
formulation of stochastic lattice models (see e.g.\ \cite{\ADHR}).
Denote the probability to find the system in a configuration
$\{ \beta \}$ by $P(\{ \beta \})$. Then the time evolution of
the system is given by the master equation
$$\partial_t P(\{ \beta \})
  = - \sum_{\{ \beta' \}} W_{\{ \beta \} \to \{ \beta' \}} P(\{ \beta \})
    + \sum_{\{ \beta' \}} W_{\{ \beta' \} \to \{ \beta \}} P(\{ \beta' \})
\label{mastereq}$$
where the two terms on the r.h.s.\ represent the loss and gain
processes respectively. The symbol $\partial_t$ can equally well
denote evolution in discrete or continuous time. To formulate
\ref{mastereq} in quantum-mechanical terms one introduces a basis of
states $\state{\{ \beta \}}$ corresponding to the configurations
$\{ \beta \}$ and specifies the state of the system by a vector
$$\state{P} = \sum_{\{ \beta \}} P(\{ \beta \}) \state{\{ \beta \}} \, .
\label{stateProb}$$
Then the master equation \ref{mastereq} takes the form of a Schr\"odinger
equation in imaginary time
$$\partial_t \state{P} = - H \state{P}
\label{schreq}$$
where the Hamiltonian $H$ for a stochastic system is in general
non-hermitean. Due to probability conservation
one always has an eigenstate of $H$ with eigenvalue zero.
\mn
In order to apply this general formalism to the \ffm{}
we choose the following basis at each site:
$$\left(\matrix{0 \cr 1 \cr}\right) = \hbox{empty place}
   = \Es \, , \qquad
\left(\matrix{1 \cr 0 \cr}\right) = \hbox{tree}
   = \Ts \, .
\label{forestBasis}$$
A general state $\state{\{ \beta \}}$ corrsponding to
a configuration of the complete system is given by
$$\state{\{ \beta \}} =
  \state{i_1 \ldots i_L} \, ,
\label{forestBasisGlob}$$
where $i_x = \Es, \Ts$ encodes the state of the site $x$.
The vectors \ref{forestBasisGlob} form a basis of a $2^L$
dimensional complex vector space.
The notation $T_x$ and $E_x$ will be used for the operators
that measure if site $x$ is occupied by a tree or empty.
\mn
We will consider the case of small $p$ and $f$. Then in each
time step changes occur essentially only at one site and one
can replace parallel dynamics by sequential dynamics.
Note that  for the original three-state model parallel
dynamics is essential because e.g.\ the probability of
fire spreading simultaneously at different places is large.
With sequential dynamics the Hamiltonian takes the form
\begineqnseries[hamOP]
$$H = p \; H_0 + f \; V
\label{hamOPform}$$
and can be written down  explicitly using Pauli matrices.
In terms of Pauli matrices the operators
$T_x$ and $E_x$ are given by $T_x = \Tpr_x$ and $E_x = \Epr_x$.
The state of a given site is changed by the operators
$\sigma^{\pm} = \oh \left( \sigma^x \pm i \sigma^y\right)$.
\mn
$H_0$ is a sum of single-site terms $(H_0)_x$ describing
growth of a tree at site $x$. In spin language, the gain term
then involves single spin flips from $\Es$ to $\Ts$:
$$H_0 = \sum_{x=1}^L \left(\Epr_x - \sip_x\right) \, .
\label{hamOPH0}$$
The term $V$ describes the burning down of tree clusters and
is given by
$$\eqalign{
V = \sum_{x=1}^L \sum_{l=1}^{L} \; l \; &  \left\{
  \Epr_{x} \Tpr_{x+1} \cdots
  \Tpr_{x+l} \Epr_{x+l+1} \right. \cr
& \left.
- \Epr_{x} \simi_{x+1} \cdots
  \simi_{x+l} \Epr_{x+l+1}  \right\} \cr
}\label{hamOPV}$$
with suitable conventions such that the projectors at
the boundary drop out for a cluster that occupies the entire
system. This term contains flips of entire clusters from
state $\Ts$ to the empty state $\Es$. The weight of such
a cluster-flip process is proportional to the size $l$ of
the cluster.
\endeqnseries
\mn
We would like to note that
our Hamiltonian \ref{hamOP} is different from the one presented
in Anhang B of \cite{\Drossel}. Also the Fock space formulations
of the master equation in \cite{\PaTri,\BoDe} are different
since they treat situations with more than two states per site.
\mn
Clearly, it is possible to choose $p=1$ in \ref{hamOPform} by
suitably rescaling $H$. Below we will assume $p=1$ and
we will be interested in small $f$. Then one can apply
standard quantum-mechanical perturbation theory to
\ref{hamOP}.
\mn
Before performing any computation one can already see that if one
wants to take the limit $L \to \infty$ of a perturbation expansion,
one is forced to approach the critical point $f=0$. This is due
to the fact that a perturbation series is expected to converge
as long as $f$ times the largest matrix element of $V$ (which
has value $L$) is small compared to the distance $1$ between the
eigenvalues of $H_0$. This amounts to imposing
$$f \; L \ll 1
\label{CONDconv}$$
in order to be on safe grounds. Thus, one
is forced to let $f \to 0$ as $L \to \infty$. In other words,
the radius of convergence of the perturbation series
is expected to be zero for $L = \infty$. Still, perturbation
theory will tell us something about the properties of the
critical point itself.
\mn
First, we use the Hamiltonian formulation of the \ffm{}
in order to determine the stationary state perturbatively.
In order to simplify the presentation
we introduce the notation
$$\state{N} :=  \sum
\state{\underbrace{\Es \Ts \ldots}_{N \, {\rm empty \, places}}}
\label{stateN}$$
for a sum over all configurations with a total of
$N$ empty places. For the stationary state $\GS$ perturbation theory
is particularly simple. One makes the power-series ansatz
$\GS = \sum_{\nu = 0}^{\infty} f^{\nu} \state{G_\nu}$.
Obviously $\state{G_0} = \state{\Ts \ldots \Ts} = \state{0}$.
For the first-order correction $\state{G_1}$, the standard
quantum-mechanical formulas lead to the condition
$H_0 \state{G_{1}} + V \state{G_0} = 0$ where the projection onto
$\state{G_0}$ has to be subtracted. Since $H_0$ as well as
$V \state{G_0}$ are invariant under permutations of the spins,
$\state{G_1}$ must also have this property and thus can
be written as a superposition of the states $\state{N}$
$$\state{G_1} = \sum_{N=1}^{L} \alpha_N \state{N} \, .
\label{expAN}$$
Let us note that, apart form the growth processes, only
one lightning process enters in this first-order calculation,
namely the one leading from the completely full system
to a completely empty system.
\mn
Using now that the coefficient of a state $\state{N}$ with
$1 \le N \le L-1$ in the expression for $H_0 \state{G_1}$ must
be zero, one finds the recurrence relation
$\alpha_{N+1} = \textfrac{N}{L-N} \alpha_N$ for $1 \le N$.
The only further condition is $\alpha_L = 1$. This gives
$\alpha_N = {L-1 \choose N-1}^{-1}$. In summary, we have found that
$$\GS = \state{\Ts \ldots \Ts} + f \;
     \sum_{N=1}^{L} {1 \over {L-1 \choose N-1}} \state{N}
         + \Order{f^2}
\label{GSpert}$$
statisfies $H \GS = \Order{f^2}$.
\mn
Eq.\ \ref{GSpert} is in the form that arises naturally from
perturbation theory, but is is not yet properly normalized
as a probablility. Thus, we multiply the state $\GS$ with
${\cal N}_L$ and impose the normalization condition that
the sum of all coefficients in the state ${\cal N}_L \GS$ is one.
This leads to
$${1 \over {\cal N}_L} = 1 + f \;
    \sum_{N=1}^{L} {{L \choose N} \over {L-1 \choose N-1}}
      + \Order{f^2}
  = 1 + f L \sum_{N=1}^L {1 \over N}  + \Order{f^2} \, .
\label{normVal}$$
{}From \ref{GSpert} one can compute every correlation function
for $f \to 0$. Let
$$n_r(s) = \langle E_{x_1} \ldots E_{x_{r+1}} T_{x_{r+2}} \ldots
              T_{x_{s+r+1}} \rangle
\label{nrsDEF}$$
be any correlation function with $s$ trees and $r+1$ empty places
at certain fixed sites. Since \ref{GSpert} is invariant under a
permutation of the $L$ sites of the lattice, this correlation
function will only depend on the total number of trees and
empty places, respectively.
{}From  \ref{GSpert} one finds that
$$n_r(s) = {\cal N}_L \; f \; \sum_{N=r+1}^{L-s}
               {{L-s-r-1 \choose N-r-1} \over {L-1 \choose N-1}}
= {{\cal N}_L \; f \; L \; r \hbox{!} \over (s+1) (s+2) \cdots (s+r+1)}
\label{nrsRES}$$
for any {\it finite} system, $r \ge 0$  and $0 \le s \le L-r-1$.
The second identity is a purely combinatorial one and proven in
appendix A for $r=0$. For $r>0$ the result follows most easily from
the equation of motion for $n_r(s)$ in the stationary state
(compare section 5). Note that any correlation function involving
at least one empty place decays algebraically in this approximation,
the chain length $L$ acting as cutoff. Furthermore,
tree-tree or cluster-cluster correlation functions are
trivial in the sense that they do not depend on the
distance between the trees, respective clusters.
\mn
Let us briefly comment on a few special cases of \ref{nrsRES}:
\item{1)} For the cluster-size distribution $n(s) = n_1(s)$
          one recovers the result eq.\ \ref{EQns} with $1-\rho =
          {\cal N}_L f L$. A derivation of this result for $r=1$
          along similar lines as above and in section 5 below can also
          be found in \cite{\CDrSchw,\CDrSchwA}.
          In fact, the discussion of the stationary state at small $f$
          in \cite{\CDrSchw} inspired us to apply quantum-mechanical
          perturbation theory to the \ffm{}.
\item{2)} The probability to find a cluster of $s$ empty
          places with two adjacent trees is given by
          $n_{s-1}(2) = \textfrac{2 {\cal N}_L f L}
          {s (s+1) (s+2)}$. The functional form of
          this correlation function was already derived
          in \cite{\BakPac}. Note that in eq.\ (5) of
          \cite{\BakPac} which was used to obtain this result,
          the loss terms by a lightning stroke at the boundaries
          are missing. This is, however, irrelevant for the
          derivation of the form of $n_{s-1}(2)$.
\item{3)} For the probability to find a string of $s$
          adjacent empty sites one finds that
          $n_{s-1}(0) = \Textfrac{{\cal N}_L f L}{s}$
          for $1 \le s \le L-1$. For this special case it is
          also comparably easy to directly prove the combinatorial
          identity \ref{nrsRES} by induction on $L$.
\item{4)} The probability $K(s)$ to find $s$ trees at certain
          places and {\it no} empty places can be obtained from
          \ref{CONnsK} using \ref{nrsRES} for $n(s)$.
          One finds in particular that $K(s) \sim \ln(s)$ for
          large $s$. Note that the smaller the exponent of
          a correlation function the sooner will it be dominated
          by an exponential decay for $f \ne 0$. In particular, the
          asymptotic logarithmic behaviour of $K(s)$ is therefore
          not observable in simulations (compare Fig.\ 3
          of section 2).
\mn
It is also straightforward to compute the mean density of trees $\rho$.
One finds that $\textfrac{\rho}{1 - \rho} \sim \ln(L)$ for $L$
large. This also shows that in this first-order approximation only the
finite size of the system prevents it from being critical.
\mn
We now proceed with a discussion of the relaxation spectrum.
For $f=0$ the operator $H = H_0$ has real eigenvalues which are
simple integers $\Lambda = 0, 1, \ldots, L$. In general, one can use
the fact that for periodic boundary conditions the lattice
translation operator commutes with the Hamiltonian and the spectrum
splits into $L$ sectors with different momenta $P$. We introduce the
notation
$$\pstate{i_1 i_2 \ldots i_{N-1} i_N}_P :=
    \state{i_1 i_2 \ldots i_{N-1} i_N }
   + e^{i P} \state{i_N i_1 i_2 \ldots }
   + \ldots + e^{i (L-1) P} \state{i_2 \ldots i_{N-1} i_N i_1}
\label{defP}$$
for states of fixed momentum $P = \Textfrac{2 \pi n}{L}$ and
specify the first Brillouin zone by $0 \le n < L$,
$n \in \Zed$.
\mn
Some of the excited states can easily be written down.
Using that a sum of all $L$th roots of unity vanishes, one can
see that
$$\eqalign{
H \pstate{\Es \Ts \ldots \Ts}_P
   =& (1 + f (L-1)) \pstate{\Es \Ts \ldots \Ts}_P \cr
H \pstate{\Es \Es \Ts \ldots \Ts}_P
   =& (2 + f (L-2)) \pstate{\Es \Es \Ts \ldots \Ts}_P
   + (1 + e^{i P}) \pstate{\Es \Ts \ldots \Ts}_P \cr
}\label{exEig}$$
for $P \ne 0$. Eq.\ \ref{exEig} implies that there are {\it exact}
excited states with eigenvalue $1 + f (L-1)$ and $2 + f (L-2)$
respectively.  However, for $L$ large they will not belong to the
low-lying part of the excitation spectrum.
\mn
Other excited states can be treated with standard quantum
mechanical perturbation theory, bearing in mind that the left
and right eigenvectors have to be treated separately.
Some computations are presented in appendix B. The final result
for the states with eigenvalue $\Lambda = 1$ of $H_0$ is that $H$
has $L-1$ times the eigenvalue $1+f(L-1)+ \Order{f^2}$ and one
times the eigenvalue $1+f(2L-1)+ \Order{f^2}$.
Similarly, the states with eigenvalue $\Lambda = 2$ of $H_0$
give rise to an eigenvalue $2+f(L-2) + \Order{f^2}$ of $H$
for each momentum (including the translationally invariant sector).
In the translationally invariant sector this eigenvalue
$2 + f(L-2) + \Order{f^2}$ can cross with the eigenvalue
$1 + f (2 L - 1) + \Order{f^2}$ around $f = \textfrac{1}{L+1}$.
The physically interesting region is located beyond this
crossing where the lowest eigenvalues form a complex-conjugate
pair. However, there are fundamental obstacles to reach
this region beyond the crossing with perturbation expansions
around $f = 0$.
\bn
\section{Numerical computation of the relaxation spectrum}
\mn
In order to obtain results for the thermodynamic limit
of the relaxation spectrum, one has to resort to a numerical
diagonalization the Hamiltonian at finite $L$. In this section we
argue on the basis of such computations that the \ffm{}
has a non-vanishing complex gap in the thermodynamic limit for
$f > 0$. The exponents describing the critical behaviour for $f \to 0$
of the real and imaginary parts of this complex gap are both
non-integral.
\mn
After projecting onto the eigenspaces \ref{defP}
with fixed momentum $P$, the Hamiltonian \ref{hamOP} becomes
a square matrix of approximate size $2^L/L$.
Up to $L = 20$, where the size of the matrix
is slightly larger than $50000$, a few extremal eigenvalues can
be obtained using standard methods.
\mn
In addition to translational invariance, the model
is invariant under parity $x \mapsto -x$. This can be exploited
to reduce the dimensionality of the problem a little further
for $P=0$ and $P=\pi$. More important are the following
consequences of the invariance with respect to parity:
\item{1)} The spectra in the sectors with momenta $P$
          and $2 \pi - P$ are equal.
\item{2)} In each sector with fixed momentum, the
          eigenvalues are either real or come in complex
          conjugate pairs. This holds because the characteristic
          polynomial of the complete Hamiltonian \ref{hamOP}
          is real, and because of property 1).
\mn
Fig.\ 4 shows the low-lying excitations $\Lambda$ for $L=20$ and
$f=0.1$ in the complex plane.
Because of the symmetries mentioned above we have restricted
ourselves to $0 \le P \le \pi$ and to ${\rm Im}(\Lambda) \ge 0$.
Note that the figure does not include the ground state $\Lambda = 0$
and that there are many more eigenvalues near its right end
and beyond.
\mn
\centerline{\psfig{figure=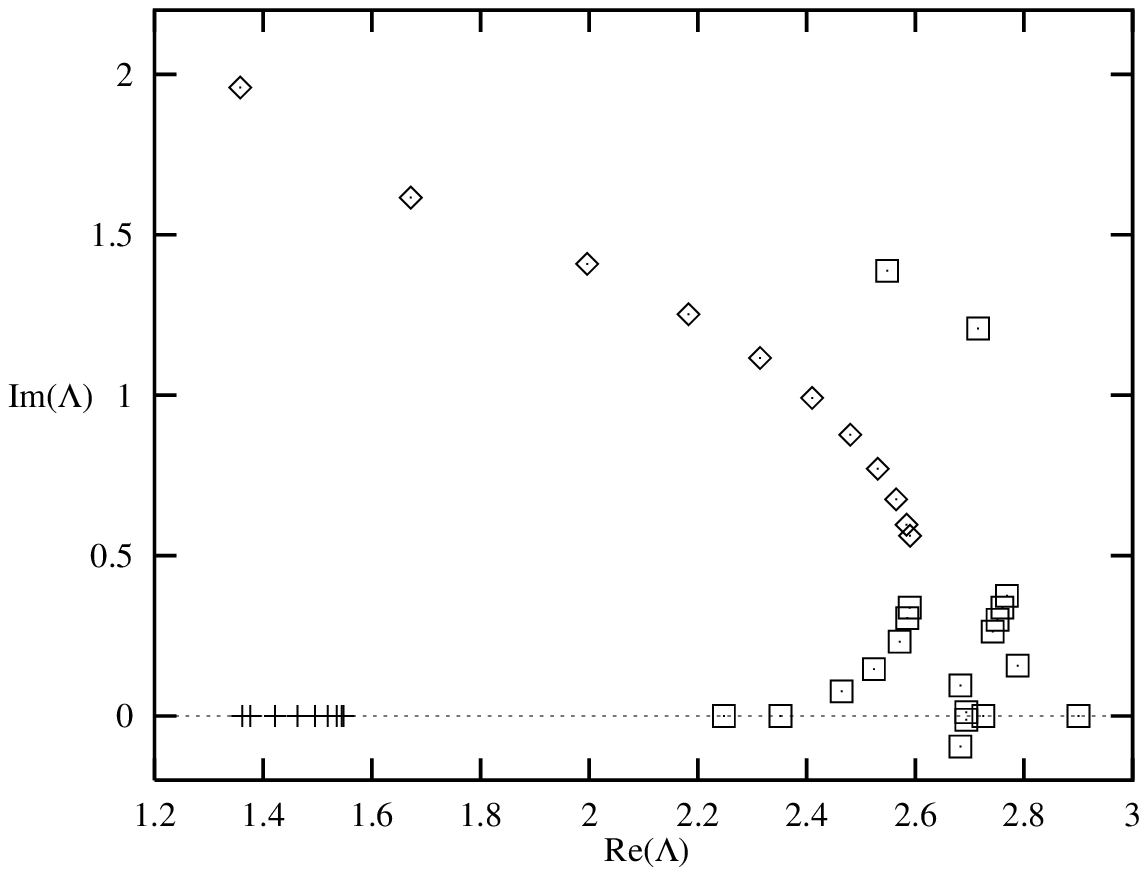}}
\sn
{\par\noindent\figindents
{\bf Fig.\ 4:} Eigenvalues of the forest-fire Hamiltonian with
               $L=20$ sites, $p=1$, $f=0.1$. For
               the symbols compare the text.
\par\noindent}
\mn
The comparably large value of $f=0.1$ has been chosen in
order to have the dispersionless modes \ref{exEig} separated
from the low-lying part of the spectrum. The first one of
these modes is located at $\Lambda = 2.9$ in
Fig.\ 4. One observes a group of levels on the real axis
indicated by the symbols `$+$'. The level with the smallest
gap has $P = \Textfrac{4 \pi}{L}$. Then the gap increases
with increasing $P$ and reaches its maximum at $P=\pi$.
One can also clearly see another sequence of complex-conjugate
levels (which are drawn with the symbol `\SymbolB')
starting around $\Lambda = 1.4 \pm 2 i$ for $P=0$ and
tending to  $\Lambda = 2.6 \pm 0.56 i$ for $P \to \pi$.
There are also further levels in this region. The ones
we have computed are marked by the symbol `\SymbolA' in Fig.\ 4.
They do not form groups as clearly as the levels described
before. It is also possible that some of them will still move
considerably with increasing $L$ due to their interaction with
the mode at $\Lambda = 1 + f (L-1)$. The figure demonstrates that
the relaxational spectrum of the Hamiltonian \ref{hamOP} does not
have a simple particle interpretation. The levels drawn with
the symbols `$+$' and `\SymbolB' are isolated levels for $P$
fixed and therefore clearly not multi-particle states which
would form continua, neither are their features typical for
single-particle states. In particular, single-particle states
usually have exponentially small corrections in the chain length,
but the levels under consideration have larger corrections
(compare also below). Even if the levels corresponding to
the symbols `$+$' and `\SymbolB' were considered as fundamental
single-particle states, some of the remaining levels which we
denoted by `\SymbolA' would remain unexplained.
\mn
In the remainder of this section we concentrate on the behaviour of
the lowest gap $\Eex$ in the translationally invariant sector. In
Fig.\ 4 one has $\Eex \approx 1.3579 \pm 1.9586 i$. Whether the
level $\Eex$ actually has the smallest real part, or if other
levels exhibit different critical behaviour would require more
detailed investigations which are beyond the scope of the present
paper. The restriction to $P=0$ is convenient
because only in the translationally invariant sector the
finite-size effects can be analyzed easily. In addition,
measurements involving spatial averages project onto this
sector, so that it plays a  particular r\^ole.
\mn
Table 1 contains asymptotic values obtained numerically for the gap
$\Eex$. They were obtained by computing numerically the
lowest complex conjugate levels
\footnote{$^{1})$}{Typically these levels fall below others
(with respect to the real part), which are located on the real axis,
for $L \sim 10$ in the range of $f$ considered here.}
for $5 \le L \le 20$ and extrapolating to $L = \infty$ using
the van den Broeck-Schwartz algorithm (see e.g.\ \cite{\SchueHen}).
\mn
\centerline{\vbox{
\hbox{
\vrule \hskip 1pt
\vbox{ \offinterlineskip
\def\tablespace{height2pt&\omit&&\omit&&\omit&&\omit&&\omit&\cr}
\def\tablerule{ \tablespace
                \noalign{\hrule}
                \tablespace        }
\hrule
\halign{&\vrule#&
  \strut\hskip 4pt\hfil#\hfil\hskip 4pt\cr
height5pt& \omit && \multispan3 && \multispan3 & \cr
& \omit && \multispan3 \hfil diagonalization ($p=1$)\hfil && \multispan3
   \hskip 4pt\hfil simulation with $p=1/200$, $L=10000$ \hfil\hskip 4pt& \cr
height5pt& \omit && \multispan3 && \multispan3 & \cr
\noalign{\hrule}
\tablespace
\tablespace
& $f/p$  &&
   $\lim\limits_{L \to \infty} \Eex$ &&
       error of $\vert\lim\limits_{L \to \infty} \Eex\vert$
 && relaxation time && oscillation period
                                          & \cr \tablespace \tablerule
& $0.1$  && $1.2947 \pm 2.0030 i$  && $0.0003$ && $140$ &&$560$&\cr\tablespace
& $0.09$ && $1.2344 \pm 1.9625 i$  && $0.0003$ && $150$ &&$575$&\cr\tablespace
& $0.08$ && $1.1726 \pm 1.9170 i$  && $0.0002$ && $160$ &&$615$&\cr\tablespace
& $0.07$ && $1.1069 \pm 1.8664 i$  && $0.0002$ && $160$ &&$695$&\cr\tablespace
& $0.06$ && $1.0360 \pm 1.8093 i$  && $0.0004$ && $190$ &&$705$&\cr\tablespace
& $0.05$ && $0.9600 \pm 1.7461 i$  && $0.0004$ && $205$ &&$715$&\cr\tablespace
& $0.04$ && $0.8767 \pm 1.6721 i$  && $0.0003$ && $255$ &&$740$&\cr\tablespace
& $0.03$ && $0.7820 \pm 1.5820 i$  && $0.0010$ && $270$ &&$775$&\cr\tablespace
& $0.02$ && $0.6780 \pm 1.4680 i$  && $0.0040$ && $300$ &&$875$&\cr\tablespace
}
\hrule}\hskip 1pt \vrule}
\hbox{Table 1: Estimates for the first translationally invariant gap of the
               forest-fire Hamiltonian.}}
}
\mn
The values in table 1 have to be considered with some care
because the maximal system size $L=20$ used for the extrapolation to
$L = \infty$ is still quite small. However, all sequences used for the
extrapolation are monotonuos in $L$. Furthermore, the deviation at
finite $L$ from the values given above is typically of the order $L^{-2}$.
Under these conditions one can usually obtain reliable estimates
for the thermodynamic limit $L = \infty$ although the estimates
in table 1 for the error of $\Eex$ may be somewhat too optimistic.
\mn
In order to perform a further check of the extrapolations
we have simulated the temporal behaviour of the average density of
trees $\rho(t)$. This quantity approaches its limiting value $\rho(\infty)$
in an oscillatory way. The simulations were performed for $p=1/200$ on a
lattice of size $L=10000$ -- a size that is much closer to the
thermodynamic limit. For one random initial condition with
$\rho(0) = {1 \over 2}$  the time evolution of $\rho(t)$ was averaged
over 100 realizations.  Then the oscillation period was determined by
looking for crossings of $\rho(t)$ at early times $t$ with the stationary
value $\rho(\infty)$ (which can be estimated from $\rho(t)$ for large $t$).
The relaxation time can be obtained by looking at the values of
local extrema of $\rho(t)$ at early times. If only one complex
pair contributes, the
relaxation time equals $1/(p \; {\rm Re}(\Eex))$ while the oscillation
period is given by $2 \pi /(p \; {\rm Im}(\Eex))$, where $\Eex$
is the normalized relaxational mode at $p=1$.
One obtains good agreement with the extrapolations in Table 1
keeping the crude method in mind which can at most be expected to
be accurate to $10\%$.
\mn
Because the extrapolations can be performed much more accurately
we used them in order to determine the behaviour of $\Eex$ as
a function of $f$.  The data in Table 1 can nicely be fitted by
$$\lim_{L \to \infty} \Eex = 3.258 \; f^{0.405} \pm 3.127 i \; f^{0.194}
  \, .
\label{ENgapScale}$$
This is consistent with a standard critical point at $f = 0$ -- like
the simulations of the correlation functions in the previous section.
We would like to point out that this critical behaviour arises
because the limits $f \to 0$ and $L \to \infty$ do not commute,
Taking $f \to 0$ first, one has $\Eex = 1$ for all $L$.
\mn
The critical exponents in \ref{ENgapScale} may still have systematic
errors due to the fact that only systems of maximal size $L=20$ were used
and that because of this restriction in the system size we could
not get closer to the critical point than $f = 0.02$. Thus, the
critical exponents for the oscillation frequency ${\rm Im}(\Eex)$ and
inverse relaxation time ${\rm Re}(\Eex)$ could be $\Textfrac{1}{5}$ and
$\Textfrac{2}{5}$, respectively.
\bn
\section{Simplified models in the symmetric space}
\mn
In this section we consider several simplified models
that deal only with the total number of trees respectively
empty places. The main motivation is that the
critical correlation functions \ref{nrsRES} follow
most easily from such a simplified model. Because this
simplified model exhibits neither the critical behaviour
\ref{CONexp} nor the gap \ref{ENgapScale} we investigate
whether it is possible to generalize the model in the
symmetric space such that it describes also some
aspects off the critical point correctly. It turns
out that the natural generalizations do not have these
properties. This shows that most of the critical exponents
of the \ffm{} seem to be due to actual
correlations while the power laws of the critical correlation
functions \ref{nrsRES} are a global effect.
\mn
The first simplified model corresponds to the first-order
expansion \ref{GSpert}. It consists in modifying
the dynamics in such a way that this first-order expansion of
the ground state becomes exact. Since only the burning
of the completely full system contributes in first
order perturbation theory, this is achieved by dropping
all lightning processes except for the completely full
system (see also \cite{\CDrSchwA}).
This modification leaves the space of states \ref{stateN}
invariant, so one can restrict oneself to this subspace.
We call it symmetric since the states are invariant under
permutations of the lattice sites. Consequently,
there is no real spatial structure for this simplified
model any more.
\mn
It is useful to replace the basis \ref{stateN} by
the following differently normalized one:
$$\tstate{N} := {\state{N} \over {L \choose N}} \, .
\label{stateNren}$$
The normalization factor in \ref{stateNren} is precisely
the number of configurations contributing to
$\state{N}$. In this basis the simplified Hamiltonian
is given by the following $(L+1) \times (L+1)$ matrix:
$$\Hs = \pmatrix{
fL & -1 &  0 & \cdots & 0 \cr
0  &  1 & -2 & \ddots & \vdots \cr
\vdots & \ddots & 2 & \ddots & 0 \cr
0 & & \ddots & \ddots & -L \cr
-f L & 0 & \cdots & 0 & L \cr
} \, .
\label{Hsimple}$$
The normalization of the states in \ref{stateNren} is
chosen such that the Hamiltonian \ref{Hsimple} is
a manifestly stochastic one, i.e.\ the columns sum to
zero. It describes a single particle hopping on a ring with
sites $0$, $\ldots$, $L$ with the following rates:
$$\horline{0}{0}{30}
\horline{32}{0}{1}
\horline{34}{0}{1}
\horline{36}{0}{1}
\horline{38}{0}{1}
\horline{40}{0}{1}
\horline{42}{0}{1}
\horline{45}{0}{30}
\point{0}{0}
\putbox{0}{3}{\ninerm 0}
\point{15}{0}
\putbox{15}{3}{\ninerm 1}
\point{30}{0}
\putbox{30}{3}{\ninerm 2}
\point{45}{0}
\putbox{45}{3}{\nineit L\ninerm\ -- 2}
\point{60}{0}
\putbox{60}{3}{\nineit L\ninerm\ -- 1}
\point{75}{0}
\putbox{75}{3}{\nineit L}
\putbox{7.5}{-3}{$1$}
\putbox{22.5}{-3}{$2$}
\putbox{52.5}{-3}{$L-1$}
\putbox{67.5}{-3}{$L$}
\putbox{7.5}{0}{\Arrow}
\putbox{22.5}{0}{\Arrow}
\putbox{52.5}{0}{\Arrow}
\putbox{67.5}{0}{\Arrow}
\putbox{7}{-3}{\QcircA}
\putbox{82}{-3}{\QcircB}
\horline{7}{-6.6}{61}
\putbox{37.5}{-4.8}{\RArrow}
\putbox{37.5}{-9.5}{$f L$}
\hskip75mm
\label{stochPsimp}$$
By construction, the state \ref{GSpert} is the
exact ground state of the simplified Hamiltonian
\ref{Hsimple}. In the basis \ref{stateNren} it reads
$$\GS = \tstate{0} + f \; L \;
     \sum_{N=1}^{L} {1 \over N}  \tstate{N} \, .
\label{GSsimple}$$
One can use this simplified model to obtain the result
\ref{nrsRES}. The equation of motion for the quantity
$n_r(s)$ (see \ref{nrsDEF}) is:
$$\partial_t n_r(s) = -(r+1) n_r(s) + s n_{r+1}(s-1)
\label{EVnrs}$$
for $0 < s < L$, $r \ge 0$. The first term describes the
loss term by tree growth at any of the $r+1$ empty
places whereas the second term is the gain term
by a tree having grown at any of the $s$ places that
are occupied by trees in the final configuration.
Note that lightnings affect only configurations
with $s=0$ or $s=L$ trees and are therefore not relevant
for the present consideration. In the stationary state,
eq.\ \ref{EVnrs} leads to the recurrence
relation $n_{r+1}(s-1) = \Textfrac{(r+1) n_r(s)}{s}$.
Together with the combinatorial proof of \ref{nrsRES} for $r=0$
presented in appendix A this proves \ref{nrsRES} for general $r$.
\mn
Now we proceed with a discussion of the excitation
spectrum of $\Hs$.
The characteristic polynomial of $\Hs$ yields the
equation
$$(f L - \lambda) \prod_{r=1}^{L} \left(r
- \lambda\right) - f L \prod_{r=1}^{L} r = 0
\label{charPol}$$
for the eigenvalues $\lambda$. A discussion of
this characteristic polynomial leads to the following
results on the relaxational spectrum
$${\rm Im}(\lambda_k) = {2 \pi k \over \ln(L)}
   \, , \quad k \in \Zed \, , \qquad \qquad
   {\rm Re}(\lambda_1) = {\pi^4 \over 3 \ln(L)^3}
\label{EVSPEC}$$
for $L$ large.
A numerical computation of the lowest eigenvalues of $\Hs$
up to $L \approx 10^{3}$ yields good agreement with the
result \ref{EVSPEC} for ${\rm Im}(\lambda_k)$. The precise
behaviour of the real part is more difficult to verify, but
from the numerical computations one sees that the real parts
of at least the lowest eigenvalues tend much faster to zero
than their imaginary parts.
\mn
The result for ${\rm Im}(\lambda)$, as well as the eigenfunctions
can also be obtained from a continuum approximation. Details
of the computation are presented in appendix C. The
continuum model shows that the eigenfunctions are essentially
plane waves in the variable
$\Textfrac{\ln(N)}{\ln(L)}$ multiplied by $\Textfrac{1}{N}$.
In this variable, the $k$th eigenfunction has $k$ periods.
Therefore, the value $\Textfrac{1}{{\rm Im}(\lambda_k)}$ can
be interpreted as the average time that a particle needs to
hop around the $k$th part (on this logarithmic scale)
of the chain.
\mn
In summary, the gap of the simplified Hamiltonian
\ref{Hsimple} vanishes as $\Textfrac{1}{\ln(L)}$ as
$L \to \infty$ irrespective of the value of $f$
\footnote{${}^{2})$}{
The logarithmic dependence of the relaxation time on the
system size is reminiscent of results obtained in \cite{\BuHe}
for certain two-dimensional kinetic spin models with
cluster dynamics.
}.
This is different from the behaviour of the full \ffm{}
as observed in the previous section.
\mn
The simplified model \ref{Hsimple} discussed so far is not
able to describe the behaviour of the full \ffm{} away from
the critical point. As a first step in this direction one can
take all loss terms due to lightning in the full model into account.
In order to have a stochastic process these loss terms must
be balanced with suitable gain terms. A convenient choice
is to attribute all gain terms to the completely empty system.
Thus, in this generalization of \ref{Hsimple} we introduce
transitions from the state $\tstate{N}$ to the state $\tstate{L}$
with weight $f (L-N)$. Then the picture \ref{stochPsimp}
turns into:
$$\horline{0}{5}{15}
\horline{17}{5}{1}
\horline{19}{5}{1}
\horline{21}{5}{1}
\horline{23}{5}{1}
\horline{25}{5}{1}
\horline{27}{5}{1}
\horline{32}{5}{1}
\horline{34}{5}{1}
\horline{36}{5}{1}
\horline{38}{5}{1}
\horline{40}{5}{1}
\horline{42}{5}{1}
\horline{45}{5}{15}
\point{0}{5}
\putbox{0}{8}{\ninerm 0}
\point{15}{5}
\putbox{15}{8}{\ninerm 1}
\point{30}{5}
\putbox{30}{8}{\nineit N}
\point{45}{5}
\putbox{45}{8}{\nineit L\ninerm\ -- 1}
\point{60}{5}
\putbox{60}{8}{\nineit L}
\putbox{7.5}{2}{$1$}
\putbox{52.5}{2}{$L$}
\putbox{7.5}{5}{\Arrow}
\putbox{52.5}{5}{\Arrow}
\verline{0}{-3}{8}
\verline{60}{-3}{8}
\putbox{7}{-6}{\QcircA}
\putbox{67}{-6}{\QcircB}
\horline{7}{-9.6}{46}
\putbox{30}{-7.8}{\RArrow}
\putbox{30}{-12.5}{$f L$}
\putbox{37}{2}{\QcircA}
\putbox{67}{2}{\QcircB}
\horline{37}{-1.6}{16}
\putbox{45}{0.2}{\RArrow}
\putbox{45}{-4.5}{$f (L-N)$}
\putbox{17}{-1.6}{$\cdots\cdots\cdots$}
\hskip60mm
\label{stochPgemp}$$
It is straightforward to obtain the ground state of the
Hamiltonian corresponding to \ref{stochPgemp} either
numerically up to $L \approx 10^4$, or from a continuum
approximation. Then one finds for the cluster-size distribution
$$n(s) \sim s^{-(2 + f L)}
\label{nsPgemp}$$
for $s \gg 1$. It is tempting to interpret this result
as a continuously varying critical exponent but then
one has to introduce a rescaled rate $\hat{f} = f L$ in
order to keep the exponent finite in the thermodynamic limit.
In any case, the loss processes by lightning strokes in
\ref{stochPgemp} do not account for the finite correlation
length $\xi_c$. This is
surprising because consideration of the same processes (with
some additional approximations) lead to the correct critical
behaviour \ref{CONexp} of the correlation length $\xi_c$
\cite{\CDrSchw}. A solution to this puzzle might be the
following: Once one assumes that there is an upper cutoff,
the loss processes by lightning strokes fix its critical
exponent although these processes alone do not explain
the presence of this cutoff.
\mn
Other possible modifications of the simplified Hamiltonian
$\Hs$ are:
\item{1)} One can introduce death of trees irrespective of
          their neighbourhood with a rate $q$, thereby
          modeling the ignition processes which dominate in
          the full model if the number of trees is small.
          For $f=0$ one then has a simple kinetic model
          where the spins flip independently and $n(s)$
          varies exponentially with $s$. However, this
          does not give an upper cutoff in the general case
          since then $n(s)$ is well approximated by a
          {\it sum} of the results for $f=0$ and $q=0$,
          respectively.
\item{2)} One can distribute the loss terms in \ref{stochPgemp}
          equally over the configurations with fewer trees.
          This amounts to introducing a transition from
          a state $\tstate{N_i}$ to a state $\tstate{N_f}$
          for $N_f > N_i$ with rate $f$. Computing the ground state
          numerically as before, one finds that this
          model behaves very similarly to that in \ref{stochPgemp}.
          In particular, its cluster-size distribution also has
          the behaviour \ref{nsPgemp}.
\mn
Since none of the aforementioned models is able to
describe the off-critical behaviour correctly,
let us finally discuss the accurate projection of the
full \ffm{} onto the symmetric subspace. This
projection can be performed as follows: Consider the matrix
element of $V$ that leads from $N_i$ empty places to $N_f$
empty places, i.e.\ a cluster of $N_f - N_i$ trees
is struck by lightning. Thus, there must have been
$L-N_f$ trees outside the cluster in the initial
state. This cluster occupies $2+N_f-N_i$ places
including the two empty places at its boundary.
This leads to a combinatorial factor
${L-2-N_f+N_i \choose L-N_f}$ in the transition
matrix element. Translational invariance yields
another factor $L$ and the weight for the process
is $N_f - N_i$. Writing this in the normalization
\ref{stateNren} leads to the matrix element
$$\atstate{N_f} V \tstate{N_i} =
 {L \; (N_f-N_i) \; {L-2-N_f+N_i \choose L-N_f} \over
 {L \choose Ni}}
\label{projMat}$$
for $N_f > N_i$. The loss term due to lightning equals
$\atstate{N} V \tstate{N} = -(L-N)$ and
$\atstate{N_f} V \tstate{N_i} = 0$ for $N_f < N_i$.
The tree growth has already been correctly accounted
for in \ref{Hsimple}. Putting this together, one obtains
the correct projection of the full Hamiltonian \ref{hamOP}
onto the symmetric subspace. The corresponding matrix is almost
triangular.
\mn
As for the other simplified models, the ground state and
thus the cluster-size distribution of this Hamiltonian can
easily be obtained from a numerical diagonalization
for rings with up to a few thousands sites.
For small $f$ one finds again $n(s) \sim s^{-(2 + f L)}$
for $s \gg 1$ and {\it no} crossover
to an exponential decay. For $f$ large, $n(s)$ decays
rapidly already for small $s$. In this case, the
$L$-dependence of $n(s)$ computed in the symmetric
subspace for small $s$ becomes small and one can
even observe qualitative agreement with simulations
of the full model. However, even in the case \ref{projMat}
the critical behaviour \ref{CONexp} for $f \to 0$ is
not reproduced. In summary, the symmetric space seems unable to
describe the critical behaviour as a function of $f$,
while it does describe the critical correlation functions
\ref{nrsRES} very nicely.
\vfill
\eject
\section{Discussion}
\mn
In this paper we have studied the one-dimensional
forest-fire model and exhibited several new non-integral
critical exponents. It remains to be seen to what
extent our results are relevant for higher dimensions
as well. Firstly, it should be checked to a better
accuracy whether $\nu = \nu_T$ \cite{\Henley} is
indeed true in two dimensions. Another question to
be addressed is if in higher dimensions the spatial
power laws associated with clusters can also be obtained
by looking at global quantities only. In particular,
it remains an open problem if models in a symmetric
subspace can also describe the critical cluster-size
distribution in higher dimensions. The two-dimensional
version of the \ffm{} introduced in \cite{\CDrSchP}
seems to point in this direction. In this case, the
dynamics is also controlled by a global quantity,
namely the mean density of trees.
\mn
The fact that the critical density in one
dimension is $\rho = 1$ leads to the unusual situation that
all correlation functions at the critical point
become independent of the spatial variables in the
thermodynamic limit. At $f=0$, the density in the
stationary state equals one for all dimensions.
Now, a necessary requirement for perturbation theory
around $f=0$ to be of any use is that the density
$\rho$ must be continuous for $f \to 0$. In one
dimension, this condition is satisfied, which
permitted us to employ perturbation theory and thus
led to the simplified model. The situation is different
in higher dimensions where the critical density is smaller
than one. Then correlation functions with non-trivial
spatial behaviour at the critical point are possible,
but perturbation theory around $f=0$ cannot be used.
\mn
Even if at present our results seem to be specific for
one dimension, our simple model in the symmetric space
shows that spatial power laws can arise from purely global
effects. A similar observation for certain models of evolution
was made recently in \cite{\BJW} where the impact of replacing
local neighbourhoods by random neighbourhoods was examined.
In our case, this global dynamics arises naturally at the
critical point and is not put in by hand. We have also shown
that this simplified model has a vanishing gap, i.e.\ physical
quantities will relax with a temporal power law to the
stationary state. This demonstrates that neither power laws in
static quantities related to avalanche type processes nor slow
relaxational modes are a sufficient criterion for criticality
in a strict sense, so that the usual multi-point correlations
should always be studied in addition.
\bn
\displayhead{Acknowledgments}
\mn
We thank B.\ Drossel, M.\ Kaulke and F.\ Schwabl
for discussions and comments.  A.H.\ would like to thank the
Deutsche Forschungsgemeinschaft for financial support.
\sectionnumstyle{Alphabetic}
\newsectionnum=0
\vfill
\eject
\appendix{A combinatorial identity}
\mn
We want to show that
$$S(L,s) := \sum_{N=1}^{L-s} {{L-s-1 \choose N-1} \over {L-1 \choose N-1}}
  = {L \, \over s+1}
\label{combIden}$$
for $0 \le s \le L-1$. The proof uses induction
on $L$. First one observes that
$$S(L,L-1) = 1
\label{indstart}$$
which gives the start of the induction. Now one can
proceed with the following induction step
$$S(L+1,s) = \sum_{N=1}^{L+1-s}
        \prod_{u=0}^{s-1} {L+1-N-u \over L-u}
         = 1 + {L-s \over L} S(L,s) \, .
\label{indstep}$$
Observing that the r.h.s.\ of \ref{combIden} satisfies
the recurrence relation \ref{indstep} completes the proof
of this combinatorial identity.
\bn
\appendix{Perturbation expansion for excitations}
\mn
In this appendix we indicate how to compute the excitation
spectrum of the \ffm{} using quantum-mechanical
perturbation theory. First we note that $H_0$ in \ref{hamOP}
can easily be diagonalized. In matrix form the part at one site
is given by
$$(H_0)_x = \pmatrix{0 & -1 \cr 0 & 1 \cr} \, .
\label{H0x}$$
The right eigenvectors of this matrix are
$$\left(\matrix{1 \cr 0\cr}\right) \, , \qquad
\left(\matrix{-1 \cr 1\cr}\right)
\label{rightEV}$$
with eigenvalues $0$ and $1$, respectively.
The corresponding left eigenvectors are
$$(1 \, 1) \, , \qquad (0 \, 1) \, .
\label{leftEV}$$
Now one can treat excited states with standard quantum-mechanical
perturbation theory where one just has to keep
in mind that the left eigenvectors \ref{leftEV} are not just the
transposed ones of the right eigenvectors \ref{rightEV}.
{}From \ref{rightEV} and \ref{leftEV} it follows immediately that
the right vectors
$$\state{x} = - \state{\Ts \ldots \Ts}
    + \state{\Ts \ldots \Ts \Esl_x \Ts \ldots \Ts}
\label{rightState}$$
and the left vectors
$$\astate{y} = \sum_{i_1, i_2 \ldots = \Es}^{\Ts}
   \astate{i_1 i_2 \ldots \Esl_y \ldots}
\label{leftState}$$
are eigenvectors of $H_0$ with eigenvalue $1$. The normalization
in \ref{rightState} and \ref{leftState} is already
$\langle y \state{x} = \delta_{y,x}$. Now one easily checks that
$$\astate{y} V \state{x} = \left[
  \pmatrix{L-1 & & \cr & \ddots& \cr & & L-1}
+ \pmatrix{1 & \cdots & 1 \cr \vdots & & \vdots \cr 1 & \cdots  & 1}
   \right]_{y \, x} \, .
\label{matEn1}$$
The eigenvalues of the matrix \ref{matEn1} are straightforward to
compute. The result is that $H$ has one eigenvalue
$1+f(2L-1)+ \Order{f^2}$ and $(L-1)$ degenerate eigenvalues
$1+f(L-1)+ \Order{f^2}$. The latter correspond to the result
\ref{exEig} for $P \ne 0$, and in this case the first-order expansion
happens to be exact for all $f > 0$. The new result of the
perturbation expansion is the eigenvalue $1+f(2L-1)+ \Order{f^2}$
which is located in the sector $P=0$.
\mn
Similarly, one finds that the $\Textfrac{L (L-1)}{2}$ vectors
$$\state{x_1, x_2} = \state{\Ts \ldots \Ts}
- \state{\ldots \Ts \Esl_{x_1} \Ts \ldots}
- \state{\ldots \Ts \Esl_{x_2} \Ts \ldots}
+ \state{\ldots \Ts \Esl_{x_1} \Ts \ldots \Ts \Esl_{x_2} \Ts \ldots}
\label{rightSt}$$
with $x_1 < x_2$ are right eigenvectors of $H_0$ with eigenvalue 2.
The corresponding left eigenvectors are given by
$$\astate{y_1, y_2} = \sum_{i_1, i_2 \ldots = \Es}^{\Ts}
    \astate{i_1 i_2 \ldots \Esl_{y_1} \ldots \Esl_{y_2} \ldots}
\label{leftSt}$$
with $y_1 < y_2$. These vectors are properly normalized, i.e.\
$\langle y_1, y_2 \state{x_1, x_2} = \delta_{y_1, x_1} \delta_{y_2, x_2}$.
A straightforward computation yields
$$\astate{y_1, y_2} V \state{x_1, x_2} =
(L-2) (1 + \delta_{y_1, x_1} \delta_{y_2, x_2})
- \cases{(x_2 - x_1 -1) & if $x_1 \le y_1 \, , y_2 \le x_2$ \cr
(L - x_2 + x_1 -1) & if $\left\{
\matrix{
\hbox{$y_1 \le x_1 \, , x_2 \le y_2$} \cr
\hbox{or $x_2 \le y_1$} \cr
\hbox{or $y_2 \le x_1$} \cr
}\right.$  \cr
 0 & otherwise. \cr
}\label{matEn2}$$
It is easy to check that the $L$ vectors $\state{x, x+1}$
are eigenvectors of \ref{matEn2} with eigenvalue $(L-2)$.
In particular, there is an eigenvalue $2+f(L-2) + \Order{f^2}$ of $H$
for each momentum (including the translationally invariant sector) which is
compatible with the result \ref{exEig} for $P \ne 0$.
\mn
All first-order corrections discussed so far are positive, but
the matrix \ref{matEn2} also has negative eigenvalues. Since
the eigenvalues of a stochastic matrix always have positive
real parts, this automatically implies that such a negative
correction can only be relevant for small $f$. More precisely,
one finds eigenvalues $2 - f h(L,P) + \Order{f^2}$ where
$h(L,P)$ is positive around $P=\pi$, symmetric in $P-\pi$
and increases for small $\abs{P - \pi}$.
This does however not affect the translationally invariant sector,
nor is it relevant for the large $L$ limit. Nevertheless, the
complicated $P$-dependence of these modes
demonstrates that the structure of the full relaxation spectrum
is highly non-trivial.
\vfill
\eject
\appendix{Continuum approximation}
\mn
The model described by $\Hs$ in section 5 can be treated in the
following continuum approximation which gives the imaginary
part of the spectrum and the eigenfunctions in a simple way.
\mn
Denote the eigenvalues by
$\lambda$ and the components of the corresponding
eigenvector by $g_0$, ..., $g_L$. Then the eigenvalue
problem for \ref{Hsimple} is given by the following equations:
\begineqnseries[EVHsimple]
$$\eqalignno{
f L g_0 - g_1 &= \lambda g_0 \, , &\eqnlabel{EVeqA} \cr
N g_N - (N+1) g_{N+1} &= \lambda g_N \, ,
  \qquad 1 < N < L \, , &\eqnlabel{EVeqB} \cr
-f L g_0 + L g_L &= \lambda g_L \, . &\eqnlabel{EVeqC} \cr
}$$
\endeqnseries
The solution of the inner equations
$-(N+1)(g_{N+1} - g_N) - g_N = \lambda g_N$ varies only slowly
for large $N$ so that we can approximate them by
$$-x {{\rm d} \over {\rm d}x} g(x) = (\lambda +1) g(x)
\label{KDGBL}$$
for $1 \le x \le L$. Keeping only the leading term in $L$ of
the boundary conditions \ref{EVeqA}, \ref{EVeqC} and eliminating
$g_0$ one finds
$$g(L) = {g(1) \over L} \, .
\label{KBDC}$$
For the stationary state these equations lead to
$g(x) = \Textfrac{1}{x}$ which corresponds exactly to the discrete
result $g_N \sim \Textfrac{1}{N}$.
\mn
The general solution of the differential equation \ref{KDGBL} is
given by
$$g(x) = x^{-(\lambda + 1)}
\label{KSOL}$$
and from the boundary condition \ref{KBDC} one finds the purely
imaginary spectrum
$$\lambda_k = {2 \pi i k \over \ln(L)} \, , \qquad k \in \Zed \, .
\label{KSPEC}$$
The eigenfunctions are plane waves in the variable
$\Textfrac{\ln(x)}{\ln(L)}$ multiplied by $\Textfrac{1}{x}$
and compare well with those of the discrete problem
\ref{EVHsimple}, except for small $x$.
\vfill
\eject
\displayhead{References}
\mn
\tolerance=10000
\bibitem{  1} P.\ Bak, C.\ Tang, K.\ Wiesenfeld, {\it Self-Organized
              Criticality: An Explanation of $1/f$ Noise}, Phys.\ Rev.\ Lett.\
              {\bf 59} (1987) 381-384
\bibitem{  2} P.\ Bak, C.\ Tang, K.\ Wiesenfeld, {\it Self-Organized
              Criticality}, Phys Rev.\ {\bf A38} (1988) 364-374
\bibitem{\BCT} P.\ Bak, K.\ Chen, C.\ Tang, {\it A Forest-Fire Model and Some
              Thoughts on Turbulence}, Phys.\ Lett.\ {\bf A147} (1990)
              297-300
\bibitem{\BakS} P.\ Bak, K.\ Sneppen, {\it Punctuated Equilibrium and
              Criticality in a Simple Model of Evolution}, Phys.\ Rev.\ Lett.\
              {\bf 71} (1993) 4083-4086
\bibitem{\FBS} P.\ Bak, H.\ Flyvbjerg, K.\ Sneppen, {\it Mean Field Theory for
              a Simple Model of Evolution}, Phys.\ Rev.\ Lett.\ {\bf 71}
              (1993) 4087-4090
\bibitem{\GrKa} P.\ Grassberger, H.\ Kantz, {\it On a Forest Fire Model with
              Supposed Self-Organized Criticality}, J.\ Stat.\ Phys.\ {\bf
              63} (1991) 685-700
\bibitem{\DMS} B.\ Drossel, W.K.\ Mo{\ss}ner, F.\ Schwabl, {\it Computer
              Simulations of the Forest-Fire Model}, Physica {\bf A190}
              (1992) 205-217
\bibitem{\DrSchwA} B.\ Drossel, F.\ Schwabl, {\it Self-Organized Critical
              Forest-Fire Model}, Phys.\ Rev.\ Lett.\ {\bf 69} (1992)
              1629-1632
\bibitem{\LPVZ} V.\ Loreto, L.\ Pietronero, A.\ Vespignani, S.\ Zapperi, {\it
              Renormalization Group Approach to the Critical Behavior of the
              Forest Fire Model}, Phys.\ Rev.\ Lett.\ {\bf 75} (1995) 465-468
\bibitem{\DrSchF} B.\ Drossel, F.\ Schwabl, {\it Self Organization in a
              Forest-Fire Model},  Fractals {\bf 1} (1993) 1022-1029
\bibitem{\CDrSchE} S.\ Clar, B.\ Drossel, F.\ Schwabl, {\it Scaling Laws and
              Simulation Results for the Self-Organized Critical Forest-Fire
              Model}, Phys.\ Rev.\ {\bf E50} (1994) 1009-1018
\bibitem{\BakPac} P.\ Bak, M.\ Paczuski, {\it Theory of the One-Dimensional
              Forest-Fire Model}, Phys.\ Rev.\ {\bf E48} (1993) R3214-R3216
\bibitem{\CDrSchw} S.\ Clar, B.\ Drossel, F.\ Schwabl, {\it Exact Results for
              the One-Dimensional Self-Organized Critical Forest-Fire Model},
              Phys.\ Rev.\ Lett.\ {\bf 71} (1993) 3739-3742
\bibitem{\Henley} C.L.\ Henley, {\it Statics of a ``Self-Organized''
              Percolation Model}, Phys.\ Rev.\ Lett.\ {\bf 71} (1993) 2741-2744
\bibitem{\ADHR} F.C.\ Alcaraz, M.\ Droz, M.\ Henkel, V.\ Rittenberg, {\it
              Reaction-Diffusion Processes, Critical Dynamics, and Quantum
              Chains}, Ann.\ Phys.\ {\bf 230} (1994) 250-302
\bibitem{\CDrSchwA} S.\ Clar, B.\ Drossel, F.\ Schwabl, {\it Universality in
              the One-Dimensional Self-Organized Critical Forest-Fire Model},
              Z.\ Naturforsch.\ {\bf 49} (1994) 856-860
\bibitem{\BJW} J.\ de Boer, A.D.\ Jackson, T.\ Wettig, {\it Criticality in
              Simple Models of Evolution}, Phys.\ Rev.\ {\bf E51} (1995)
              1059-1074
\bibitem{\Grassb} P.\ Grassberger, {\it On a Self-Organized Critical
              Forest-Fire Model},
              J.\ Phys.\ A: Math.\ Gen.\ {\bf 26} (1993) 2081-2089
\bibitem{\FYWu} F.Y.\ Wu, {\it The Potts Model}, Rev.\ Mod.\ Phys.\ {\bf 54}
              (1982) 235-268
\bibitem{\GJS} G.\ Grinstein, C.\ Jayaprakash, J.E.S.\ Socolar, {\it On
              Self-Organized Criticality in Nonconserving Systems}, Phys.\
              Rev.\ {\bf E47} (1993) 2366-2376
\bibitem{\Drossel} B.\ Drossel, {\it Strukturbildung in offenen Systemen},
              Ph.D.\ thesis, M\"unchen (1994)
\bibitem{\PaTri} H.\ Patzlaff, S.\ Trimper, {\it Analytical Approach to the
              Forest-Fire Model}, Phys.\ Lett.\ {\bf A189} (1994) 187-192
\bibitem{\BoDe} S.\ Bottani, B.\ Delamotte, {\it Towards a Field Theory
              Analysis of Self-Organized Criticality: The Forest Fire and
              Sandpile Models}, LPTHE preprint 95/10
\bibitem{\SchueHen} M.\ Henkel, G.M.\ Sch\"utz, {\it Finite-Lattice
              Extrapolation Algorithms}, J.\ Phys.\ A: Math.\ Gen.\ {\bf 21}
              (1988) 2617-2633
\bibitem{\BuHe} A.N.\ Burkitt, D.W.\ Heermann, {\it System Size Dependence of
              the Autocorrelation Time for the Swendsen-Wang Ising Model},
              Physica {\bf A162} (1990) 210-214
\bibitem{\CDrSchP} S.\ Clar, B.\ Drossel, F.\ Schwabl, {\it Self-Organized
              Critical and Synchronized States in a Nonequilibrium
              Percolation Model}, Phys.\ Rev.\ Lett.\ {\bf 75} (1995)
              2722-2725
\vfill
\end